\documentclass[%
 reprint,
nofootinbib,
 amsmath,amssymb,
 aps,
]{revtex4-1}

\usepackage{graphicx}
\usepackage{dcolumn}
\usepackage{bm}
\usepackage{multirow}
\usepackage[usenames]{color}

\newcommand{\la}{\langle}
\newcommand{\ra}{\rangle}

\newcommand{\dbar}{\bar{d}}

\begin{document}

\preprint{APS/123-QED}

\title{\boldmath Rates and Asymmetries of $B\to\pi\,\ell^+\ell^-$ Decays}

\author{
Wei-Shu Hou, 
Masaya Kohda, 
and Fanrong Xu 
}
\affiliation{
Department of Physics, National Taiwan University, Taipei, Taiwan 10617 
}


\begin{abstract}
The LHCb experiment observed $B^+ \to \pi^+ \mu^+ \mu^-$ decay
with $1.0~\mathrm{fb}^{-1}$ data, which is the first measurement of
a flavor changing neutral current
$b\to d\ell^+\ell^-$ decay ($\ell=e, \mu$).
Based on QCD factorization, we give Standard Model predictions
for the branching ratios, direct CP asymmetries,
and isospin asymmetry for $B\to\pi\ell^+\ell^-$ decays,
in the kinematic region where the dilepton invariant mass is small.
We find that the contribution from weak annihilation enhances
the direct CP asymmetry for low $\ell^+\ell^-$ pair mass.
Anticipating improved measurements,
we assess the utility of $B^+\to\pi^+\ell^+\ell^-$ observables,
when combined with $B^0 \to \pi^-\ell^+\nu$ and $B^+\to K^+\ell^+\ell^-$,
for determining CKM parameters in the future.

\begin{description}
\item[PACS numbers]
12.15.Hh 
11.30.Er 
13.20.He 
\end{description}
\end{abstract}

\pacs{Valid PACS appear here}
\maketitle



\section{\label{sec:Intro}INTRODUCTION\protect\\}

After much anticipation for possible new physics
in $b\to s$ transitions, all measurements at the
LHC with 2011--2012 data turned out consistent with
Standard Model (SM) expectations, and
further progress would take more 
 data to unfold.
The focus, and perhaps sensitivity to New Physics,
may now be on $b\to d$ rare decays.

The LHCb has observed \cite{LHCb:2012de}
the $B^+\to \pi^+\mu^+\mu^-$ decay with measured branching ratio
\begin{align}
&\mathcal B(B^+\to \pi^+\mu^+\mu^-) \notag\\
&= ( 2.3\pm0.6\; ({\rm stat.}) \pm0.1\;({\rm syst.}) ) \times 10^{-8},
\label{eq:LHCb}
\end{align}
at $5.2\,\sigma$ significance.
This is the first observation of a flavor changing neutral current (FCNC)
$b\to d\ell^+\ell^-$ transition,
as the B factories were only able to set
limits~\cite{Wei:2008nv,Lees:2013lvs}.
We summarize relevant data in Table~\ref{tab:exp}.
As data accumulates, the era of FCNC $b\to d$ rare decays has dawned.
The Belle II experiment under construction
should be able to measure $B^0\to \pi^0\ell^+\ell^-$ decay in the future.

Previous theoretical studies are based on
naive factorization~\cite{Aliev:1998sk,Song:2008zzc,Wang:2007sp,Ali:2013zfa}
or the perturbative QCD approach~\cite{Wang:2012ab}.
In this paper, we provide a first estimate for
the exclusive $B\to \pi \ell^+\ell^-$ ($\ell=e$ or $\mu$)
decays based on the QCD factorization (QCDF)
framework~\cite{Beneke:1999br,Beneke:2000ry,Beneke:2000wa,
Beneke:2001at,Ali:2001ez,Bosch:2001gv,Beneke:2004dp}.
This framework has been introduced for exclusive nonleptonic $B$-meson decays~
\cite{Beneke:1999br,Beneke:2000ry}, and applied for
exclusive radiative and semileptonic $B$-meson FCNC decays~
\cite{Beneke:2000wa,Beneke:2001at,Ali:2001ez,Bosch:2001gv,Beneke:2004dp},
which include the well-studied $B\to K^*\ell^+\ell^-$.
We extend the known result for the $B\to \rho \ell^+\ell^-$ decays~\cite{Beneke:2004dp}
to the $B\to \pi \ell^+\ell^-$ case.
We provide the SM predictions for the decay rates within QCDF,
as well as associated CP asymmetries and isospin asymmetry at $\mathcal O(\alpha_s)$.

In contrast to naive factorization treatments,
one advantage of the QCDF framework is that
one can include hard-spectator-scattering effects,
such as weak annihilation diagrams.
This is crucial for addressing CP violation (CPV) and
isospin asymmetry
\footnote{
As for $B\to K^{(*)}\ell^+\ell^-/\rho\ell^+\ell^-$, 
isospin asymmetries were also studied \cite{Lyon:2013gba} 
employing light-cone sum rule to calculate matrix elements for 
weak annihilation and 
$\mathcal O_8$ [in Eq. (\ref{eq:operators})] \cite{Dimou:2012un}.
}
in these decays, which are of great interest even within SM
because $V^*_{ud}V_{ub}$ carries CPV phase.
In contrast, such effects are suppressed by
$V^*_{us}V_{ub}$ for $b\to s\ell^+\ell^-$ processes.
The $B\to \pi\ell^+\ell^-$ modes may also be used to probe
New Physics, such as two Higgs doublet model~\cite{Aliev:1998sk,Song:2008zzc},
R-parity violating supersymmetry~\cite{Wang:2007sp}, and
fourth generation~\cite{Hou:2013btm}.

This paper is a followup of the last paper,
where some results were used in conjunction
with the study of $B_d \to \mu^+\mu^-$ mode.
In this paper we focus on SM expectations for $B\to \pi\ell^+\ell^-$.
After presenting the basic formulas and input parameters in Sec. II,
the SM predictions are given in Sec. III.
We discuss prospects for constraining CKM parameters in Sec. IV,
before giving our conclusion.
Some details are placed in Appendices.

\begin{table}[b]
\caption{
A summary of experimental results for
the $B\to \pi \ell^+\ell^-$ decay
branching ratios (in unit of $10^{-8}$).
The values quoted for Belle (based on 657M $B\bar B$ pairs)
and BaBar (based on 471M $B\bar B$ pairs) are
the 90\% C.L. upper limits.
The less stringent limit by Belle on $B^0\to \pi^0 \ell^+\ell^-$
has more to do with a large central value
(for both $\mu^+\mu^-$ and $e^+e^-$),
than a somewhat larger error than BaBar.
}
\begin{center}
\begin{tabular}{l|ccc}
\hline\hline
Mode  & LHCb~\cite{LHCb:2012de} & Belle~\cite{Wei:2008nv} & BaBar~\cite{Lees:2013lvs} \\
\hline
$B^+\to\pi^+\mu^+\mu^-$ & $2.3\pm 0.6\pm 0.1$ & $< 6.9$ & $< 5.5$ \\
$B^+\to\pi^+e^+e^-$ & --  & $< 8.0$ & $< 12.5$ \\
$B^+\to\pi^+\ell^+\ell^-$    & -- & $< 4.9$ & $< 6.6$ \\
 & & & \\
$B^0\to\pi^0\mu^+\mu^-$ & -- & $< 18.4$ & $< 6.9$ \\
$B^0\to\pi^0e^+e^-$         & -- & $< 22.7$ & $< 8.4$ \\
$B^0\to\pi^0\ell^+\ell^-$   & -- & $< 15.4$ & $< 5.3$ \\
\hline\hline
\end{tabular}
\end{center}
\label{tab:exp}
\end{table}%
%


\section{\label{sec:Formula}BASIC FORMULAS and input}

The effective Hamiltonian for $b\to d\ell^+\ell^-$
is given by
\begin{align}
\mathcal H_{\rm eff}
&= -\frac{G_F}{\sqrt{2}}\left[
 \lambda_t \mathcal H_{\rm eff}^{(t)}
 +\lambda_u \mathcal H_{\rm eff}^{(u)} \right]
 +\text{h.c.},
\label{eq:Heff}
\end{align}
in SM, where $\lambda_q = V_{qd}^*V_{qb}$ ($q=u, t$) and
\begin{align}
\mathcal H_{\rm eff}^{(u)}
 &\equiv C_1\left( \mathcal{O}_1^c -\mathcal{O}_1^u \right) +C_2\left( \mathcal{O}_2^c
 -\mathcal{O}_2^u \right), \notag\\
\mathcal H_{\rm eff}^{(t)}
 &\equiv C_1\mathcal{O}_1^c +C_2\mathcal{O}_2^c
 +\sum_{i=3}^{10} C_i\mathcal{O}_i.
\end{align}
We follow the operator basis of Ref.~\cite{Beneke:2004dp}, i.e.
\begin{align}
 &\mathcal{O}_7
  =-\frac{e \hat m_b}{8\pi^2} \dbar\sigma^{\mu\nu}(1+\gamma_5) b F_{\mu\nu}, \notag\\
 &\mathcal{O}_8
  =-\frac{g_s\hat m_b}{8\pi^2}\dbar_i \sigma^{\mu\nu}(1+\gamma_5) T^A_{ij}b_j
   G^A_{\mu\nu}, \notag\\
 &\mathcal{O}_{9}
  =\frac{\alpha}{2\pi}\left[ \dbar\gamma^\mu(1-\gamma_5) b \right]
   \left[ \bar \ell \gamma_\mu\ell \right], \notag\\
 &\mathcal{O}_{10}
  =\frac{\alpha}{2\pi}\left[ \dbar\gamma^\mu(1-\gamma_5) b \right]
   \left[ \bar \ell \gamma_\mu\gamma_5\ell \right],
\label{eq:operators}
\end{align}
where repeated indices are summed,
$\alpha = e^2/4\pi$ and $\hat m_b(\mu)$ denotes the $b$ quark mass in the $\overline{\rm MS}$ scheme.
We have eliminated $\lambda_c=V_{cd}^*V_{cb}$ in Eq.~(\ref{eq:Heff}) by using
the unitarity relation $\lambda_u+\lambda_c+\lambda_t=0$.

The effective Hamiltonian for $b\to s\ell^+\ell^-$ can be obtained by replacing
$d$ by $s$ in the equations above.
In this case, the $\lambda_u \mathcal H_{\rm eff}^{(u)}$ term
can be neglected due to smallness of
$|V_{us}^*V_{ub}| \ll |V_{ts}^*V_{tb}|$.
For $b\to s\ell^+\ell^-$, therefore, the effective Hamiltonian can be factorized by
the single CKM factor $V_{ts}^*V_{tb}$ to good approximation.
In contrast,
$\lambda_u$ and $\lambda_t$ are comparable in magnitude for $b \to d\ell^+\ell^-$,
with the sizable phase difference,
\begin{align}
\phi_2 \equiv \alpha \equiv
\arg \left( -\frac{V_{td}V_{tb}^*}{V_{ud}V_{ub}^*} \right),
\end{align}
for which global analyses give $\phi_2 \sim 89^\circ$~\cite{PDG}.
As $\lambda_u$ enters the amplitudes with tree-level operators
$\mathcal O_2^q = [\bar d\gamma^\mu(1-\gamma_5)q][\bar q\gamma_\mu(1-\gamma_5)b]$
($q=u,c$), its effects can be numerically large.
Hence, the amplitudes for $b\to d\ell^+\ell^-$ have more complex structure
than the $b\to s\ell^+\ell^-$ amplitudes,
resulting in richer phenomenology such as CPV,
which will be discussed in this article.
The following formulas can also be applied to
exclusive $B \to K\ell^+\ell^-$ decays in a straightforward manner,
keeping the above remarks in mind.

We use next-to-next-to-leading logarithmic (NNLL)
results for $C_{9,10}$ and next-to-leading logarithmic (NLL)
results for $C_{1-6}$ and $C^{\rm eff}_{7,8}$,
which are necessary to calculate $b\to d\ell^+\ell^-$ at $\mathcal O(\alpha_s)$.
These were obtained by using the formal solutions of the renormalization group equations
given in Ref.~\cite{Beneke:2001at},
with two-loop matching conditions~\cite{Bobeth:1999mk} and
the three-loop anomalous dimension matrix~
\cite{Chetyrkin:1996vx,Gambino:2003zm,Gorbahn:2004my},
as well as updated input parameters.

In calculating the amplitude for the exclusive decay $\bar B \to P\ell^+\ell^-$,
where $P$ denotes light pseudoscalar mesons $\pi^-$, $\pi^0$, $K^-$ and $\bar K^0$,
one has to deal with matrix elements
for the various operators appearing in Eq.~(\ref{eq:Heff}).
The matrix elements for $\mathcal O_{9,10}$ are simply expressed
by the form factors, as they are bilinear in the quark fields.
The contributions from remaining operators enter through the process
with virtual photon $\gamma^*$, namely,
$\bar B \to P \gamma^* \to P \ell^+\ell^-$,
which can be parameterized as
\begin{align}
&c_P\,\langle \gamma^*(q,\mu)P(p^\prime)| \mathcal H_{\rm eff}^{(i)} |\bar B(p) \rangle \notag\\
&= -\frac{em_b}{4\pi^2}\frac{\mathcal{T}_P^{(i)}(q^2)}{M_B}
 \left[q^2(p^\mu+p^{\prime\mu})-(M_B^2-m_P^2)q^\mu\right],
\end{align}
for $i=t,u$, where $|P \rangle$ denotes  $|\pi^- \rangle$ or $|K^- \rangle$
($|\pi^0 \rangle$ or $|\bar K^0 \rangle$) for $B^-$ ($\bar B^0$) decay.
The isospin factor $c_P$ is $-\sqrt{2}$ for $P=\pi^0$,
and $1$ for $P=\pi^-, K^-, \bar K^0$
which differs in normalization from Ref.~\cite{Beneke:2004dp}.

In the QCDF framework, the $\bar B \to P \gamma^*$ amplitude
can be decomposed as~\cite{Beneke:2001at, Beneke:2004dp}
\begin{align}
&\mathcal T_P^{(i)}
= \xi_P C_P^{(i)} \notag\\
 &+\frac{\pi^2 f_Bf_P}{N_c M_B} \sum_{\pm}
 \int_0^\infty \frac{d\omega}{\omega}\Phi_{B,\pm}(\omega)\int_0^{1}du\, \phi_P(u)
 T_{P,\pm}^{(i)}(u,\omega),
\label{QCDF-B2Pgam}
\end{align}
where $C_P^{(i)}$ and $T_{P,\pm}^{(i)}$ are described by short-distance physics,
their perturbative expressions are given in Appendix~\ref{app:formula}.
Information from long-distance physics is
encoded in the $\bar B\to P$ form factor $\xi_P$,
and the light-cone distribution amplitudes,
$\Phi_{B,\pm}$ for the $B$ meson and
$\phi_P$ for the light pseudoscalar meson.

Eq.~(\ref{QCDF-B2Pgam}) relies on the heavy quark limit of the $b$ quark.
It further assumes the energy of the final state meson $P$
in the $B$-meson rest frame,
denoted as $E_P = (M_B^2+m_P^2-q^2)/2M_B$, is large enough,
i.e. $E_P \simeq M_B/2$ or  $q^2 \ll M_B^2$.
%
The QCDF approach is, hence, restricted in the kinematical region
where the invariant mass of the lepton pair $q^2$ is small.
Besides this, the $\bar B \to P \gamma^*$ amplitude suffers from
nonperturbative corrections due to the near-threshold $u\bar u$ and $c\bar c$
intermediate states, which form the $\rho, \omega$, \ldots and charmonium resonances.
For better theoretical control, we limit our analysis
in the $2$~GeV$^2 < q^2 < 6$~GeV$^2$ region
where the QCDF approach is expected to work,
while the resonance contaminations would
also be avoided.\footnote{
The resonance effects on the $B\to\pi\ell^+\ell^-$ decays
were studied based on naive factorization in Ref.~\cite{Aliev:1998sk}.
Their result on the dilepton invariant mass distribution of the branching ratio
and of the CP asymmetry supports safety of this $q^2$ range.
The recent study~\cite{RQM} in relativistic quark model, however,
seems to suggest that resonance effects may still be
important, even for our conservative $q^2$ range.
}
The choice of $q^2$ range would be, however, rather subjective, 
given the lack of actual data and model-independent studies 
\footnote{
As for $B\to K^{(*)}\ell^+\ell^-$ decays,
a model-independent analysis on nonperturbative effects can be found in 
Ref.~\cite{Khodjamirian:2010vf}.
}
on $B\to \pi\ell^+\ell^-$ around the resonant regions.
Indeed, choices of wider $q^2$ ranges can be found in literature, e.g., 
$1$~GeV$^2 < q^2 < 6$~GeV$^2$ adopted in a study~\cite{Beneke:2004dp} 
of the closely related process $B\to \rho\ell^+\ell^-$ based on QCDF, 
or $1$~GeV$^2< q^2 < 8$~GeV$^2$ adopted as the low-$q^2$ region 
in a study~\cite{Ali:2013zfa} of $B^\pm \to \pi^\pm\ell^+\ell^-$ based on naive factorization
with the heavy quark limit.
Given this situation, we will also provide our numerical results for these two regions,
in addition to the conservative range $2$~GeV$^2 < q^2 < 6$~GeV$^2$.

Including the contributions from $\mathcal O_{9,10}$, the amplitude for
$\bar B\to P \ell^+\ell^-$ is given by
\begin{align}
&\mathcal{M}(\bar B\to P \ell^+\ell^-) \notag \\
&= \frac{G_F\alpha}{2\sqrt{2}\pi}c_P^{-1} \xi_P
 \left[\left(\lambda_t \mathcal{C}_{9,P}^{(t)}
 +\lambda_u \mathcal{C}_{9,P}^{(u)}\right)(p^\mu +p^{\prime\mu})
 (\bar{\ell}\gamma_\mu \ell)\right. \notag\\
 &\quad\quad\quad\quad\quad\quad\ \
  +\lambda_t C_{10} (p^\mu +p^{\prime\mu}) (\bar{\ell}\gamma_\mu\gamma_5 \ell)
\Big],
\end{align}
where
\begin{align}
\mathcal{C}_{9,P}^{(t)}(q^2)
&= C_9+\frac{2m_b}{M_B}\frac{\mathcal{T}^{(t)}_P(q^2)}{\xi_P(q^2)}, \notag\\
\mathcal{C}_{9,P}^{(u)}(q^2)
&= \frac{2m_b}{M_B}\frac{\mathcal{T}^{(u)}_P(q^2)}{\xi_P(q^2)}.
\end{align}
In the above amplitude, the term proportional to $m_{\ell}$ is neglected.
Therefore, hereafter we do not distinguish the cases of $\ell=e$ and $\mu$.

\begin{table*}[t!]
\caption{
Summary of input parameters,
taken from the Particle Data Group~\cite{PDG}
unless otherwise stated.
We use three-loop running for QCD coupling $\alpha_s$
with the listed initial value $\alpha_s(M_Z)$.
Our treatment of $m_b$ follows Ref.~\cite{Beneke:2001at,Beneke:2004dp}
by choosing the potential-subtracted (PS) mass~\cite{Beneke:1998rk} as input.
We define $\mathcal B_{\pi\ell\nu}^{\rm exp}
=\mathcal{B}(B^0\to \pi^- \ell^+\nu)_{q^2<12\;\rm{GeV}^2}^{\rm exp}$ as described
in the text.
The errors explicitly shown are taken into account in our error analysis.
}
\begin{center}
\begin{tabular}{c|c||c|c}
\hline\hline
$\alpha$  & $1/137$
 & $\lambda$ & $0.22535\pm 0.00065$ \\
$\sin^2\theta_W$  & $0.23$
 & $A$ & $0.811^{+0.022}_{-0.012}$ \\
$G_F$ & $1.166~\times 10^{-5}~\rm{GeV}^{-2}$
 & $\bar\rho$ & $0.131^{+0.026}_{-0.013}$ \\
$\alpha_s(M_Z)$ & $0.1184\pm0.0007$
 & $\bar\eta$ & $0.345^{+0.013}_{-0.014}$ \\
$M_W$ & $80.4$~GeV
 & $\xi_\pi (0)$ & $0.26^{+0.04}_{-0.03}$~\cite{Duplancic:2008ix}  \\
$M_Z$ & $91.2$~GeV
 & $\alpha_{\rm BK}$ & $0.53\pm 0.06$~\cite{Duplancic:2008ix} \\
$m_{t, \rm pole}$ & $173.5$~GeV
 & $a_2^\pi$ &  $0.25\pm 0.15$~\cite{Ball:2006wn}  \\
$m_{b, \rm PS}(2~\rm GeV)$ & $(4.6\pm 0.1)$~GeV~\cite{Beneke:1998rk}
 & $a_4^\pi$ & $-a_2^\pi + (0.1\pm 0.1)$~\cite{Ball:2004ye}  \\
$m_{c, \rm pole}$ & $1.67$~GeV
 &  $\xi_K (0)$ & $0.36^{+0.05}_{-0.04}$~\cite{Duplancic:2008tk} \\
$f_{\pi}$ & $(130.41\pm 0.03\pm0.20)$~MeV
 & $a_1^K$ &  $0.10\pm 0.04$~\cite{Chetyrkin:2007vm} \\
$f_{K}$ & $(156.1\pm 0.2\pm0.8\pm 0.2)$~MeV
 & $a_2^K$ &  $0.25\pm 0.15$~\cite{Ball:2006wn,Duplancic:2008tk}  \\
$f_B$ & $(190.6\pm 4.7)$~MeV~\cite{Laiho:2009eu}
 & $\xi_K (0)/\xi_\pi(0)$ & $1.38^{+0.11}_{-0.10}$~\cite{Duplancic:2008tk} \\
$\tau_{B^0}$ & $1.52\times 10^{-12}$~s
 & $\lambda_{B,+}(1.5~\rm{GeV})$ & $(0.485\pm0.115)$~GeV
 ~\cite{Braun:2003wx,Beneke:2004dp} \\
$\tau_{B^\pm}$ & $1.64\times 10^{-12}$~s
 & $\mathcal B_{\pi\ell\nu}^{\rm exp}$
 & $(0.81\pm 0.02\pm 0.03) \times 10^{-4}$~\cite{Amhis:2012bh} \\
\hline\hline
\end{tabular}
\end{center}
\label{tab:input-para}
\end{table*}
\begin{table*}[t!]
\caption[]{The SM Wilson coefficients at the scale $\mu=4.6\ \mathrm{GeV}$
in leading logarithmic (LL), next-to-leading logarithmic (NLL) and
next-to-next-to-leading logarithmic order (NNLL).
Input parameters listed in Table~\ref{tab:input-para} are used.
}
{
$$
\begin{array}{l|cccccccccc}
\hline\hline
& C_1 & C_2 & C_3 & C_4 & C_5 & C_6 & C_7^{\mathrm{eff}} & C_8^{\mathrm{eff}} & C_9 & C_{10}\\
\hline
\mathrm{LL} & -0.5093 & 1.0256 & -0.0050 & -0.0686 & 0.0005 & 0.0010 & -0.3189 & -0.1505 & 2.0111 & 0 \\
\hline
\mathrm{NLL} & -0.3001  & 1.0080 & -0.0047 &-0.0827  & 0.0003 & 0.0009 & -0.2969 & -0.1642 & 4.1869  & -4.3973 \\
\hline
\mathrm{NNLL} & - & - & - & - & - & - & - & - & 4.2607 & -4.2453 \\
\hline
\hline
\end{array}
$$
}
\label{tab:WC}
\end{table*}

The kinematics of $\bar B \to P \ell^+\ell^-$ decay can be described by
the dilepton invariant mass $q^2$ and $\cos\theta$,
where $\theta$ is the angle between the momentum of $\ell^+$ and
the momentum of $P$ in the rest frame of the lepton pair.
The $\cos\theta$ dependence of the decay might be interesting in relation to
the forward-backward asymmetry. However, it is zero in the current case
as the double differential decay rate behaves as
$d^2\Gamma/dq^2d(\cos\theta)\propto (1-\cos^2\theta)$.
Therefore, we discuss only $q^2$ dependence.

The {
 differential branching ratio} is given by
\begin{align}
&\frac{d\mathcal B}{dq^2 }(\bar B \to P \ell^+\ell^-) \notag\\
&= S_P\tau_B \frac{G_F^2M_B^3}{96\pi^3 }\left(\frac{\alpha}{4\pi}\right)^2
 \lambda(q^2,m_P^2)^3\xi_P(q^2)^2 \notag\\
&\quad \times\left(
 \left\vert\lambda_t \mathcal{C}_{9,P}^{(t)}(q^2) +\lambda_u\mathcal{C}_{9,P}^{(u)}(q^2) \right\vert^2
 +|\lambda_t|^2C_{10}^2
 \right) 
 \notag \\
&= S_P\tau_B \frac{G_F^2M_B^3}{96\pi^3 }\left(\frac{\alpha}{4\pi}\right)^2
 \lambda(q^2,m_P^2)^3\xi_P(q^2)^2 |\lambda_t|^2 \notag\\
&\quad \times\left(
 \left| \mathcal{C}_{9,P}^{(t)}(q^2) -R_{ut}e^{i\phi_2}\mathcal{C}_{9,P}^{(u)}(q^2) \right|^2
 +C_{10}^2
 \right),
\label{eq:decayrate-2}
\end{align}
where
\begin{align}
&\lambda(q^2, m_P^2) \notag\\
&=\left[\left(1-\frac{q^2}{M_B^2}\right)^2-\frac{2m_P^2}{M_B^2}\left(1+\frac{q^2}{M_B^2}\right)
+\frac{m_P^4}{M_B^4}\right]^{\frac{1}{2}}.
\end{align}
$S_P = 1/c_P^2 = 1/2$ for $P=\pi^0$, and
$S_P=1$ for $P=\pi^-, K^-, \bar K^0$, and
we have defined
\begin{equation}
\lambda_u/\lambda_t \equiv -R_{ut}e^{i\phi_2}
\label{eq:Rut}
\end{equation}
for $\bar B\to \pi\ell^+\ell^-$ in SM.

The branching ratio for the CP-conjugate mode is obtained by flipping the sign of
the weak phase in Eq.~(\ref{eq:decayrate-2}), i.e.,
by replacing $\phi_2 \to -\phi_2$. 
We then define the $q^2$-dependent direct CP asymmetries
as\footnote{
Our definitions of the CP asymmetries in Eqs.~(\ref{def:ACP}) and (\ref{def:ACP-ave})
are similar to the corresponding ones for
$B^+\to K^+\mu^+\mu^-$~\cite{Aaij:2013dgw} and
$B^0\to K^{*0}\mu^+\mu^-$~\cite{LHCb:2012kz} studied by the LHCb experiment.
}
\begin{align}
&A_{\rm CP}^+(q^2) 
 \notag\\
\equiv\;& \frac{d\mathcal B(B^-\to \pi^-\ell\ell)/dq^2
  -d\mathcal B(B^+\to \pi^+\ell\ell)/dq^2}
  {d\mathcal B(B^-\to \pi^-\ell\ell)/dq^2
  +d\mathcal B(B^+\to \pi^+\ell\ell)/dq^2}, \notag\\
&A_{\rm CP}^0(q^2) 
 \notag\\
\equiv\;& \frac{d\mathcal B(\bar B^0\to \pi^0\ell\ell)/dq^2
  -d\mathcal B(B^0\to  \pi^0\ell\ell)/dq^2}
  {d\mathcal B(\bar B^0 \to \pi^0\ell\ell)/dq^2
  +d\mathcal B(B^0\to \pi^0\ell\ell)/dq^2}.
\label{def:ACP}
\end{align}
We also define the $q^2$-dependent isospin asymmetry
as\footnote{
Our definition of the isospin asymmetry in Eqs.~(\ref{def:AI}) and (\ref{def:AI-ave})
is similar to the one used by Belle \cite{Belle-AI} and BaBar \cite{BaBar-AI},
as well as the theoretical paper \cite{Beneke:2004dp} for $B\to \rho\gamma$.
We differ from the LHCb convention \cite{LHCb-AI}
used for $B\to K^{(*)} \mu^+\mu^-$.}
\begin{equation}
A_{\rm I}(q^2)
 \equiv \frac{\tau_{B^0}}{2\tau_{B^\pm}}
 \frac{d\overline{\mathcal B}(B^+\to \pi^+\ell\ell)/dq^2}
  {d\overline{\mathcal B}(B^0\to \pi^0\ell\ell)/dq^2}-1, \label{def:AI}
\end{equation}
where $\overline{\mathcal B}$ stand for taking the
CP-average.

By integrating over the numerator and the denominator in Eq.~(\ref{def:ACP}) separately,
we further define the $q^2$-averaged direct CP asymmetries as
\begin{align}
&\langle A_{\rm CP}^+\rangle 
 \notag\\
\equiv\;& \frac{\mathcal B(B^-\to \pi^-\ell\ell)
  -\mathcal B(B^+\to \pi^+\ell\ell)}
  {\mathcal B(B^-\to \pi^-\ell\ell)
  +\mathcal B(B^+\to \pi^+\ell\ell)}, \notag\\
&\langle A_{\rm CP}^0\rangle 
 \notag\\
\equiv\;& \frac{\mathcal B(\bar B^0\to \pi^0\ell\ell)
  -\mathcal B(B^0\to  \pi^0\ell\ell)}
  {\mathcal B(\bar B^0 \to \pi^0\ell\ell)
  +\mathcal B(B^0\to \pi^0\ell\ell)},
\label{def:ACP-ave}
\end{align}
where the branching ratios $\mathcal B$ are obtained by
integrating Eq.~(\ref{eq:decayrate-2}) over a certain $q^2$ range.
Similarly, we define the $q^2$-averaged isospin asymmetry as,
\begin{equation}
\langle A_{\rm I}\rangle
 \equiv \frac{\tau_{B^0}}{2\tau_{B^\pm}}
 \frac{\overline{\mathcal B}(B^+\to \pi^+\ell\ell)}
  {\overline{\mathcal B}(B^0\to \pi^0\ell\ell)}-1. \label{def:AI-ave}
\end{equation}

With input parameters as summarized in Table~\ref{tab:input-para},
the numerical values for the Wilson coefficients at the scale $\mu=m_b$
are given in Table~\ref{tab:WC}.
Let us briefly explain some of the input parameters in Table~\ref{tab:input-para}.

There are three form factors relevant to $\bar B\to P\ell^+\ell^-$ in SM,
which are usually denoted as $f_+(q^2)$, $f_0(q^2)$ and $f_T(q^2)$.
In the heavy quark and large recoil energy (or small $q^2$) limit,
it is known~\cite{Beneke:2000wa,Charles:1998dr}
that the three form factors are related by symmetry,
and are described by the single soft form factor $\xi_P(q^2)$.
The symmetry relations are broken by $\mathcal O(\alpha_s)$ corrections,
but one may choose the factorization scheme where
$f_{+}(q^2) = \xi_P(q^2)$ holds to all orders in perturbation theory.
The $\mathcal O(\alpha_s)$ corrections to other two form factors
can be estimated perturbatively~\cite{Beneke:2000wa},
leading to 
the factorizable corrections,
$C_P^{(f,t)}$ and $T_{P,+}^{(f,t)}$,
given in Eq.~(\ref{eq:amp-t}) of Appendix A.

\begin{table*}[t!]
\renewcommand{\arraystretch}{1.2}\addtolength{\arraycolsep}{3pt}
\caption[]{
Numerical values and breakdowns for the amplitudes $\mathcal{C}_{9,P}^{(t,u)}$
at $q^2=2$~GeV$^2$
and $5$~GeV$^2$, for $B \to\pi \ell\ell$
(Table 5 in Ref.~\cite{Beneke:2004dp} gives analogous values
for $B\to\rho\ell\ell$ at $q^2=5$~GeV$^2$).
Each term is classified into two categories:
(1) form factor term, which includes $C_9$, $Y(q^2)$,
$aC_7^{\mathrm{eff}}\equiv (2m_b/M_B) C_7^{\mathrm{eff}}$,
and the $\mathcal O(\alpha_s)$ correction $C^{(1)}$;
(2) hard-spectator-scattering term, which includes
weak annihilation $T^{(0)}$
 (with the main source of strong phase underlined)
and $\mathcal O(\alpha_s)$ hard-gluon-exchange $T^{(1)}$ terms.
The ``sum'' represents the numerical values of $\mathcal{C}_{9,P}^{(t,u)}$ themselves.
Following the argument of Ref.~\cite{Beneke:2004dp} for $C_{9,\parallel}^{(t,u)}$,
we do not include $1/m_b$ corrections to the second category.
See Appendix~\ref{app:formula} for details.
}
$$
\begin{array}{c|cc|cc}
\hline\hline
\multirow{2}{*}{} &
\multicolumn{2}{c|}{q^2=2~\mathrm{GeV}^2} &
\multicolumn{2}{c}{q^2=5~\mathrm{GeV}^2} \\
\cline{2-5}
  & \mathcal{C}_{9,P}^{(t)} & \mathcal{C}_{9,P}^{(u)} &  \mathcal{C}_{9,P}^{(t)} & \mathcal{C}_{9,P}^{(u)} \\
   \hline
 C_9 & 4.26  & 0 & 4.26  & 0   \\
 Y(q^2)& 0.38 + 0.06i    & -0.50-0.85i  & 0.44 + 0.06i    & -0.17-0.85i  \\
 aC_7^{\mathrm{eff}} & -0.52   & 0   & -0.52   & 0\\
 C^{(1)} & -0.24+0.01i   & 0.24+0.77i  & -0.27-0.01i   & 0.03+0.66i   \\
 \hline
 T^{(0)} ~(\pi^-) & 0.03-0.08i   & 1.06-\underline{2.58i} & 0.03-0.02i   & 1.01-0.67i   \\
   \hspace{0.6cm}     ~(\pi^0)  & -0.02+0.04i   & 0.11-0.26i & -0.02+0.01i   & 0.10-0.07i   \\
  T^{(1)} ~(\pi^-) & 0.03-0.01 i  &  -0.08-0.03i & 0.02-0.01 i  &  -0.04-0.02i   \\
  \hspace{0.6cm}     ~(\pi^0)  & 0.01-0.01i   & -0.05+0.00i & 0.01-0.01 i  &  -0.04-0.01i  \\
  \hline
 \mathrm{sum} ~(\pi^-)  &3.95-0.06i   & 0.73-2.69i &3.97+0.03i   & 0.84-0.88i   \\
    \hspace{0.6cm}   ~(\pi^0) &3.87+0.10i   & -0.20-0.34i &3.91+0.05i   & -0.08-0.26i    \\
\hline
\hline
\end{array}
$$\label{tab:C9}
\end{table*}
%

For the $\bar B\to \pi$ form factor $\xi_\pi(q^2)$,
we adopt the above factorization scheme with
the following fit formula of Ref.~\cite{Duplancic:2008ix}
\footnote{
This has been updated in Ref.~\cite{Khodjamirian:2011ub}, where 
the low-$q^2$ result is found to be numerically consistent with Ref.~\cite{Duplancic:2008ix}.
},
obtained by QCD light-cone sum rule with aid of
the measured $q^2$-distribution of $B\to \pi\ell\nu$,
\begin{equation}
\xi_\pi(q^2)=\frac{\xi_\pi(0)}{(1-q^2/m_{B^*}^2)(1-\alpha_{\rm BK}q^2/m_B^2)}.
\end{equation}
The numerical values for the normalization $\xi_\pi(0)$ and
the slope parameter $\alpha_{\rm BK}$ are given in Table~\ref{tab:input-para}.
The light-cone distribution amplitude $\phi_\pi$ is given by
\begin{align}
\phi_\pi(u)
&=6u(1-u)\Big[ 1 +a_2^\pi\, C_2^{(3/2)}(2u-1) \notag\\
&\quad +a_4^\pi\, C_4^{(3/2)}(2u-1) +\cdots \Big],
\end{align}
where $C_n^{(3/2)}(x)$ are Gegenbauer polynomials.
For the sake of consistency, we use the same numerical values as 
Ref.~\cite{Duplancic:2008ix},
adopted to calculate the $\bar B\to \pi$ form factors,
for the Gegenbauer coefficients $a_2^\pi$~\cite{Ball:2006wn} and 
the combination $a_2^\pi+a_4^\pi$~\cite{Ball:2004ye}, 
as listed in Table~\ref{tab:input-para}, 
by neglecting $a_{n>4}^\pi$ and scale dependence.

For $B$ meson light-cone distribution amplitudes,
we follow Ref.~\cite{Beneke:2001at} and
adopt simple model functions~\cite{Grozin:1996pq},
\begin{align}
\Phi_{B,+}(\omega)
 = \frac{\omega}{\omega_0^2}e^{-\omega/\omega_0}, \quad
\Phi_{B,-}(\omega)
 = \frac{1}{\omega_0}e^{-\omega/\omega_0}.
\label{eq:BLDA}
\end{align}
These enter in the decay amplitude through the moments
$\lambda_{B,+}^{-1}$ and $\lambda_{B,-}^{-1}(q^2)$, for which the model gives
\begin{align}
&\lambda_{B,+}^{-1}
 = \int_0^\infty d\omega \frac{\Phi_{B,+}(\omega)}{\omega} =\omega_0^{-1},
\label{eq:lam+}\\
&\lambda_{B,-}^{-1}(q^2)
 = \int_0^\infty d\omega \frac{\Phi_{B,-}(\omega)}{\omega-q^2/M_B-i\epsilon}\notag\\
&~~~\quad\qquad
=\frac{e^{-q^2/(M_B\omega_0)}}{\omega_0}[-{\rm Ei}\left(q^2/M_B\omega_0\right) +i\pi ],
\label{eq:lam-}
\end{align}
where Ei$(z)$ is the exponential integral function.
We choose $\lambda_{B,+}$ as input, with numerical value
obtained by QCD sum rule calculation~\cite{Braun:2003wx,Beneke:2004dp}.
$\lambda_{B,-}^{-1}$ appears via the weak annihilation term, and
the imaginary part in Eq.~(\ref{eq:lam-}) serves
as a crucial \emph{source of strong phase}
which is necessary for CPV.
The model functions introduce theoretical uncertainty,
which is discussed in Appendix~\ref{app:model}.

\section{\label{sec:predictions} \boldmath
SM predictions \protect\\}

With the input parameters in Table~\ref{tab:input-para} and the SM Wilson coefficients
in Table~\ref{tab:WC}, we give predictions for $B\to\pi\ell^+\ell^-$.
In Table~\ref{tab:C9}, the numerical values of the amplitudes
$\mathcal{C}_{9,P}^{(t,u)}$ and their breakdowns
at $q^2=2$~GeV$^2$ and $5$~GeV$^2$ are shown.
We note that $\mathcal{C}_{9,P}^{(t)}$ is dominated by $C_9$,
which arises from the electroweak penguin and $W$-box diagrams.
For $B^0 \to \pi^0\ell^+\ell^-$, the magnitude of $\mathcal{C}_{9,P}^{(u)}$
is minor compared to $\mathcal{C}_{9,P}^{(t)}$.
On the other hand, for $B^+ \to \pi^+\ell^+\ell^-$, the magnitude of
$\mathcal{C}_{9,P}^{(u)}$ becomes comparable to $\mathcal{C}_{9,P}^{(t)}$
at $q^2=2$~GeV$^2$.
This is mainly due to the aforementioned
large imaginary part coming from
the $\mathcal O(\alpha_s^0)$ hard-spectator-scattering term $T^{(0)}$,
corresponding to the term that includes $\hat T_{P,-}^{(0,u)}$ in Eq.~(\ref{eq:amp-u}),
where $\lambda_{B,-}^{-1}$ supplies the source of the imaginary part.

\begin{figure}[b!]
\vspace{-4mm}
{\includegraphics[width=70mm]{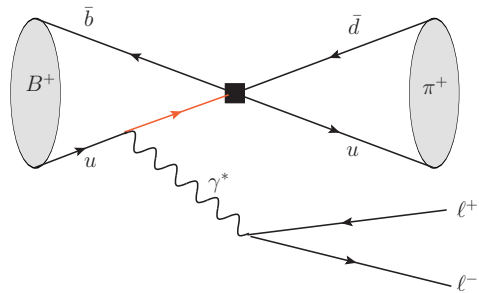}
}
\vskip0.2cm
\caption{
Tree-level weak annihilation process,
$\bar bu \to W^* \to \bar d u$ occurring inside the $B^+$ meson,
generating a large imaginary part through
the on-shell $u$ quark (in red).
} \label{fig:WeakAnnih}
\end{figure}
\begin{table*}[t!]
\renewcommand{\arraystretch}{1.2}\addtolength{\arraycolsep}{3pt}
\caption[]{
Integrated branching ratios of $B\to\pi\ell\ell$ in unit of $10^{-8}$
for $2~\rm{GeV}^2 < q^2 < 6~\rm{GeV}^2$,
obtained in two different ways:
``original'' definition using Eq.~(\ref{eq:decayrate-2});
``improved'' formula of Eq.~(\ref{eq:decayrate-imp}),
by taking the ratio with the $B\to\pi\ell\nu$ rate.
The scale uncertainty (denoted with subscript $\mu$)
is estimated by varying the scale $\mu \in [m_b/2, 2m_b]$.
}
{
$$
\begin{array}{l|c|c}
\hline\hline
 & \mathrm{Original} & \mathrm{Improved} \\
\hline
B^+\to\pi^+\ell^+\ell^- & 0.44^{+ 0.03}_{-0.02}\big|_{\mathrm{CKM}} {} ^{+0.13}_{-0.10}\big|_{\mathrm{had.}}
{}^{+0.02}_{-0.01}\big|_{\mu} & 0.47^{+ 0.05}_{-0.03}\big|_{\mathrm{CKM}} {} ^{+0.01}_{-0.01}\big|_{\mathrm{had.}}
{} ^{+0.02}_{-0.01}\big|_{\mu} {} ^{+0.02}_{-0.02}\big|_{\mathrm{\pi\ell\nu}}  \\
B^-\to\pi^-\ell^+\ell^- & 0.34 ^{+ 0.03}_{-0.02}\big|_{\mathrm{CKM}}{} ^{+0.11}_{-0.08}\big|_{\mathrm{had.}}
{}^{+0.02}_{-0.02}\big|_{\mu}
& 0.36 ^{+ 0.04}_{-0.03}\big|_{\mathrm{CKM}}{} ^{+0.01}_{-0.01}\big|_{\mathrm{had.}}
{} ^{+0.02}_{-0.02}\big|_{\mu} {} ^{+0.02}_{-0.02}\big|_{\mathrm{\pi\ell\nu}} \\
B^0\to\pi^0\ell^+\ell^- & 0.18 ^{+ 0.01}_{-0.01}\big|_{\mathrm{CKM}}{} ^{+0.06}_{-0.04}\big|_{\mathrm{had.}}
{}^{+0.01}_{-0.01}\big|_{\mu}
& 0.19 ^{+ 0.02}_{-0.01}\big|_{\mathrm{CKM}}{} ^{+0.00}_{-0.00}\big|_{\mathrm{had.}}
{} ^{+0.01}_{-0.01}\big|_{\mu} {} ^{+0.01}_{-0.01}\big|_{\mathrm{\pi\ell\nu}} \\
\bar B^0 \to\pi^0\ell^+\ell^- & 0.17 ^{+0.01}_{-0.01}\big|_{\mathrm{CKM}}{} ^{+0.05}_{-0.04}\big|_{\mathrm{had.}}
{}^{+0.01}_{-0.01}\big|_{\mu}
 & 0.18 ^{+0.02}_{-0.01}\big|_{\mathrm{CKM}}{} ^{+0.00}_{-0.00}\big|_{\mathrm{had.}}
{} ^{+0.01}_{-0.01}\big|_{\mu} {} ^{+0.01}_{-0.01}\big|_{\mathrm{\pi\ell\nu}}\\
\hline\hline
\end{array}
$$
}
\label{tab:BR}
\end{table*}

This last effect arises from the matrix element of the operator $\mathcal O_2^u$,
which is generated by the tree-level weak annihilation process
$\bar bu \to W^* \to \bar d u$ occurring inside the $B^+$ meson,
which is illustrated in Fig.~\ref{fig:WeakAnnih}.
At leading order in $1/m_b$, only the diagram where the photon is emitted by
the spectator $u$-quark in $B^+$ contributes,
and the on-shell intermediate $u$-quark, preferring smaller $q^2$,
is responsible for the imaginary part in $\lambda_{B,-}^{-1}$.
Because of the tree-level nature, this can be numerically large,
as can be read from the large Wilson coefficient $C_2 \sim 1$ in Table~\ref{tab:WC}.
At larger $q^2$, $\lambda_{B,-}^{-1}$ given in Eq.~(\ref{eq:lam-}) becomes suppressed,
and the imaginary part becomes smaller.
But the on-shell $q\bar q$ ($q=u,d,s$) intermediate states can still offer nonzero
imaginary parts at sufficiently large $q^2$, as can be read from
the imaginary part of $Y(q^2)$ [more precisely $Y^{(u)}(q^2)$]
and $C^{(1)}$ at $q^2=5$~GeV$^2$ in Table~\ref{tab:C9}.

\begin{figure*}[t!]
{\includegraphics[width=70mm]{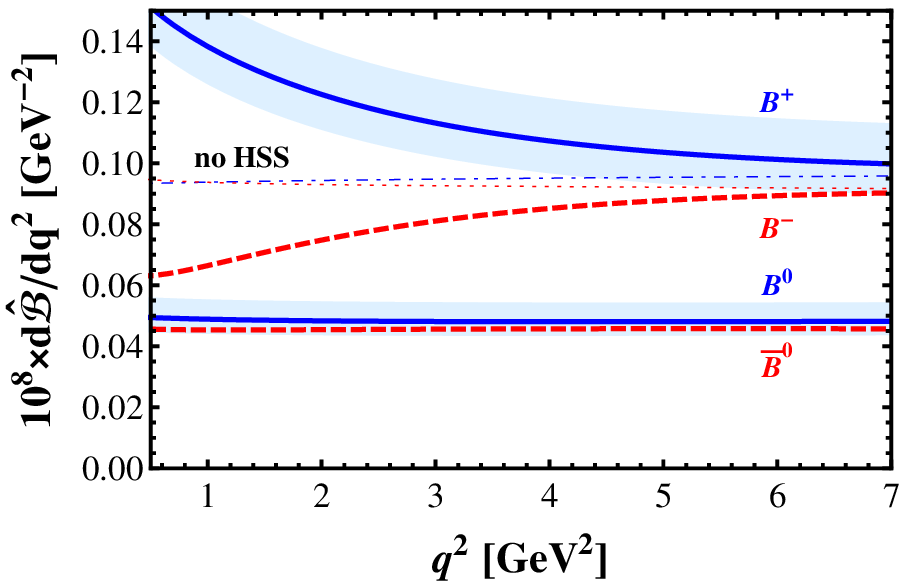} \hskip0.5cm
 \includegraphics[width=70mm]{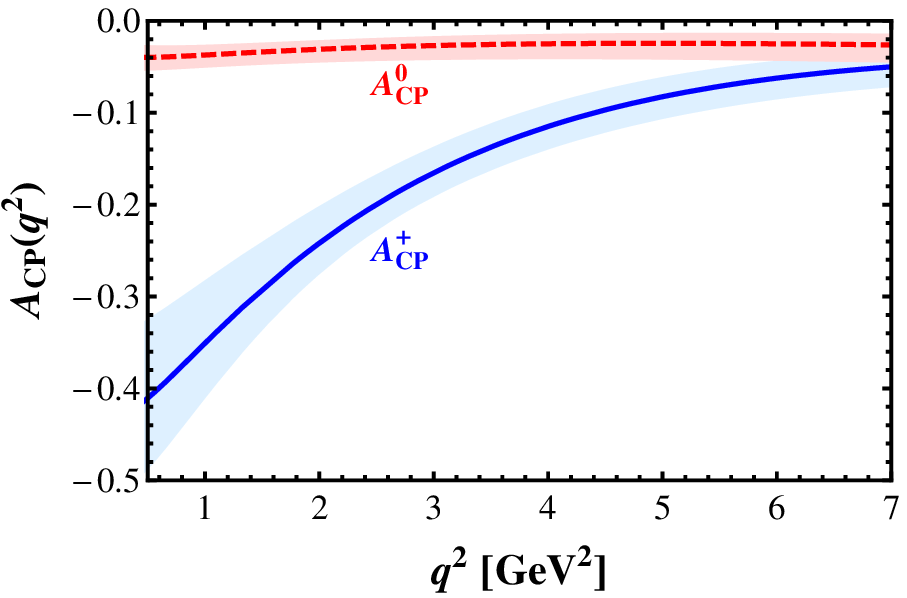}
}
\vskip-0.35cm
\caption{
Differential branching ratios $d \hat{\mathcal B}/dq^2$ [left]
and direct CP asymmetries $A_{\rm CP}(q^2)$ [right]
for $B^\pm \to \pi^\pm \ell^+\ell^-$ and $\bar B^0(B^0) \to \pi^0 \ell^+\ell^-$,
where shaded bands denote theoretical uncertainty.
In the left panel, the blue-solid (red-dashed) curves describe
$B^+$ or $B^0$ ($B^-$ or $\bar B^0$) decay,
and ``no HSS" indicates results without hard-spectator-scattering terms,
where blue dot-dashed (red dotted) curve is for $B^+$ ($B^-$).
The improved formula of Eq.~(\ref{eq:decayrate-imp}),
which takes the ratio with the $B\to\pi\ell\nu$ rate, is used.
} \label{fig:B-dif}
\end{figure*}

Using the central values and errors for input parameters in Table~\ref{tab:input-para},
we obtain a prediction for the integrated branching ratio of $B^+\to\pi^+\ell^+\ell^-$ as
\begin{align}
&\int_{2~{\rm GeV}^2}^{6~{\rm GeV}^2} dq^2
  \frac{d \mathcal B(B^+\to \pi^+\ell^+\ell^-)}{dq^2} \notag\\
&= \left( 0.44^{+ 0.03}_{-0.02}\big|_{\rm CKM} {} ^{+0.13}_{-0.10}\big|_{\rm had.}
 {}^{+0.02}_{-0.01}\big|_{\mu} \right) \times 10^{-8} \notag\\
&= \left( 0.44^{+ 0.13}_{-0.10} \right) \times 10^{-8},
\label{eq:BR-num}
\end{align}
where the dominant errors from CKM parameters, hadronic parameters and
scale uncertainty are added in quadrature.
The integration range of $2~{\rm GeV}^2 < q^2 < 6~{\rm GeV}^2$,
is for better theoretical control,
as explained in previous section.
This result and the predictions for the other three decay modes are
given in Table~\ref{tab:BR}.

The theoretical error is dominated by hadronic uncertainty, which
is in turn dominated by uncertainty in form factor normalization
$\xi_\pi(0)$.
However, this can be largely removed by taking the ratio with
the well-measured $B \to \pi \ell \nu$ rate, given by
\begin{align}
&\mathcal B_{\pi\ell\nu}
\equiv
\mathcal B (B^0 \to \pi^- \ell^+ \nu_\ell) \notag\\
&= \frac{\tau_{B^0}G_F^2|V_{ub}|^2M_B^3}{192\pi^3} \int_{q_i^2}^{q_f^2}dq^2
 \lambda(q^2,m_{\pi^-}^2)^3 \xi_\pi(q^2)^2.
\end{align}
One then obtains the improved decay spectrum,
\begin{align}
\frac{d\hat{\mathcal B}(\bar B\to \pi\ell^+\ell^-)}{dq^2}
\equiv \frac{d\mathcal B(\bar B\to \pi\ell^+\ell^-)/dq^2}{\mathcal B_{\pi\ell\nu}}
 \mathcal B_{\pi\ell\nu}^{\rm exp}.
\label{eq:decayrate-imp}
\end{align}
Given the limited range of validity for QCD light-cone sum rule calculations of the form factor,
following Ref.~\cite{Duplancic:2008ix},
we take $0 < q^2 < 12$~GeV$^2$ as the integration range for the $B\to \pi\ell\nu$ rate,
and adopt the corresponding HFAG average~\cite{Amhis:2012bh},
\begin{align}
\mathcal B_{\pi\ell\nu}^{\rm exp}
&\equiv \mathcal B(B^0\to \pi^-\ell^+\nu_\ell)^{\rm exp}_{q^2 <12~{\rm GeV}^2} \notag\\
&=(0.81 \pm 0.02 |_{\rm stat.} \pm 0.03 |_{\rm syst.} )\times 10^{-4}.
\end{align}

Using Eq.~(\ref{eq:decayrate-imp}), we then
obtain the improved integrated branching ratio of $B^+\to\pi^+\ell^+\ell^-$ as
\begin{align}
&\int_{2~{\rm GeV}^2}^{6~{\rm GeV}^2} dq^2
  \frac{d \hat{\mathcal B}(B^+\to \pi^+\ell^+\ell^-)}{dq^2} \notag\\
&= \left( 0.47^{+ 0.05}_{-0.03}\big|_{\rm CKM} {} ^{+0.01}_{-0.01}\big|_{\rm had.}
 {}^{+0.02}_{-0.01}\big|_{\mu} {}^{+0.02}_{-0.02}\big|_{\pi\ell\nu} \right) \times 10^{-8} \notag\\
&= \left( 0.47^{+ 0.06}_{-0.04} \right) \times 10^{-8}.
\label{eq:BR-num-imp}
\end{align}
Indeed, the hadronic uncertainty is reduced considerably
by cancelation of the form factor normalization $\xi_\pi(0)$,
at the cost of introducing an extra but moderate error from
$\mathcal B_{\pi\ell\nu}^{\rm exp}$ (denoted by subscript $\pi\ell\nu$).
Also, the error from CKM parameters
gets slightly enhanced due to $|V_{ub}|^2$,
which is brought in with the $B\to \pi\ell\nu$ rate.
As a whole, the total error is reduced down to 10\% level from the 20--30\%
error in the original case.

The central value of Eq.~(\ref{eq:BR-num-imp}) is slightly raised
from Eq.~(\ref{eq:BR-num}).
Given our integration range of 2--6~GeV$^2$
is smaller than the full
$0 
 \lesssim q^2 \lesssim 
 26.4$~GeV$^2$
range by more than a factor six,
it is comforting to see that our value is
smaller than the LHCb value [Eq.~(\ref{eq:LHCb})] by a factor five.
Our prediction should be checked by
LHCb with full 2011--2012 dataset,
as well as Run II and future data.

The improved predictions for all four decay modes are
also given in Table~\ref{tab:BR}.
The $B^0\to\pi^0\ell^+\ell^-$ rate is roughly a factor of
two smaller than $B^+\to\pi^+\ell^+\ell^-$, mainly due to
the isospin factor $S_{\pi^0}/S_{\pi^+}=1/2$.
Results for $q^2\in (1,6)~{\rm GeV}^2$ and
$(1,8)~{\rm GeV}^2$ 
will be given later for comparison with literature.

The $q^2$ spectra of the four $B\to\pi\ell^+\ell^-$ modes
are shown in the left panel of Fig.~\ref{fig:B-dif}.
The decay spectrum for $B^0 \to \pi^0\ell^+\ell^-$
is almost flat in the range shown, with small difference between
the CP-conjugate modes.
In contrast, the decay spectra for $B^\pm \to\pi^\pm \ell^+\ell^-$
show visible $q^2$ dependence.
At high $q^2$, the $B^+$ and $B^-$ differential rates are similar,
and larger than $B^0$ case by roughly a factor two due to isospin.
However,
at low $q^2$, the $B^+ \to\pi^+ \ell^+\ell^-$ rate tends larger
while $B^- \to\pi^- \ell^+\ell^-$ tends smaller,
which signals direct CP asymmetry.


For comparison, we also give in the left panel of Fig.~\ref{fig:B-dif} 
the $B^+$ and $B^-$ differential rates 
with the hard-spectator-scattering terms removed (indicated as ``no HSS'').
Similar to the prediction in Ref.~\cite{Ali:2013zfa},
the two decay spectra are almost degenerate and rather flat. 
This shows that hard-spectator-scattering, especially weak annihilation, 
is the main source for difference between $B^+$ and $B^-$ differential rates,
hence direct CP asymmetry, in low $q^2$ region.
On the other hand, if CP average is taken between 
the $B^+$ and $B^-$ differential rates predicted by QCDF,
the resultant decay spectrum would look quite similar to the ``no HSS'' case,
as can be read from the same figure.
In other words, the effects from the hard-spectator-scattering terms
are not so significant in the CP-averaged $B^\pm$ differential rate,
leading to a qualitatively similar prediction as naive factorization cases.
This is encouraging given the fact that a naive factorization result~\cite{Ali:2013zfa}
is already in good agreement with the LHCb result for the branching ratio in full $q^2$ range 
[Eq.~(\ref{eq:LHCb})], 
which should be understood as CP average of $B^+$ and $B^-$ decays. 
Discrimination between two predictions should require measurements
for non-CP-averaged $q^2$ spectra, or direct CP asymmetry, at low $q^2$,
with much more data.

The $q^2$-dependence of the direct CP asymmetries for
$B^{+,0}\to\pi^{+,0}\ell^+\ell^-$
are given in the right panel of Fig.~\ref{fig:B-dif}.
Indeed, the strength of $A_{\rm CP}^+(q^2)$ for $B^+\to\pi^+\ell^+\ell^-$
gets larger at lower $q^2$, reaching $-40$\%,
while $A_{\rm CP}^0(q^2)$ for $B^0\to\pi^0\ell^+\ell^-$ stays almost flat
between $-2$\% to $-4$\%.
Although we imposed $q^2 > 2$~GeV$^2$ for better theoretical control,
$A_{\rm CP}^+(q^2)$ 
can still reach $-25$\%.
For sake of showing the trend,
we have plotted both $d \hat{\mathcal B}/dq^2$ and
$A_{\rm CP}(q^2)$ in Fig.~\ref{fig:B-dif}
beyond our conservative range of 2 to 6 GeV$^2$.
%


We note that the CP asymmetry is proportional to
${\rm Im}(\lambda_t^*\lambda_u){\rm Im}({\mathcal C}_{9,P}^{(t)*}{\mathcal C}_{9,P}^{(u)})$,
where ${\rm Im}(\lambda_t^*\lambda_u)\propto \sin\phi_2$.
Besides nonzero ``weak phase'' $\phi_2$,
we also need nonvanishing phase between
${\mathcal C}_{9,P}^{(u)}$ and ${\mathcal C}_{9,P}^{(t)}$,
i.e. ``strong phase'', for generating $A_{\rm CP}$.
The large imaginary part of ${\mathcal C}_{9,P}^{(u)}$
comes from the weak annihilation term (see Fig.~\ref{fig:WeakAnnih}),
especially for low $q^2$,
with subdominant effect from real $q\bar q$ intermediate states,
which can be read from Table~\ref{tab:C9}.
On the other hand, ${\mathcal C}_{9,P}^{(t)}$ is dominated by
the real valued $C_9$ even for low $q^2$.
This results in net strong phase difference
and therefore large CP asymmetry at low $q^2$.

\begin{figure}[t!]
{
 \includegraphics[width=70mm]{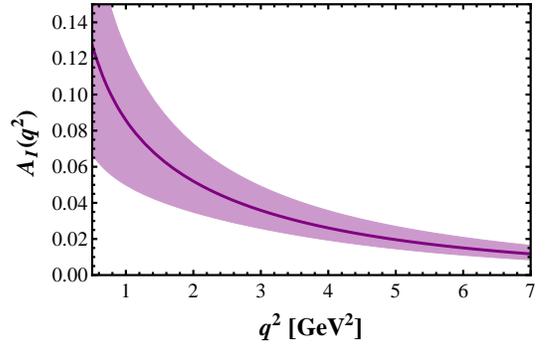}
}
\vskip-0.15cm
\caption{
Isospin asymmetry $A_{\rm I}(q^2)$ for $B\to \pi\ell^+\ell^-$ decays,
with shaded area indicating theoretical errors.
} \label{fig:asym}
\end{figure}

The CP asymmetries averaged over our $q^2$ range
are
\begin{align}
\la A_{\rm CP}^+\ra
&= -0.13^{+0.01}_{-0.01}\;\big|_{\mathrm{CKM}}\,{}^{+0.02}_{-0.02}\big|_{\mathrm{had.}}
\,{}^{+0.01}_{-0.02}\big|_{\mu} \notag\\
&=-0.13^{+0.02}_{-0.03}, \\
\la A_{\rm CP}^0\ra
&=-0.03^{+0.00}_{-0.00}\;\big|_{\mathrm{CKM}}\,{}^{+0.00}_{-0.00}\big|_{\mathrm{had.}}
\,{}^{+0.01}_{-0.02}\big|_{\mu} \notag\\
&=-0.03^{+0.01}_{-0.02},
\end{align}
which can be obtained from Table~\ref{tab:BR} directly.
Because of averaging over a $q^2$ range, the rather
large CP asymmetry at low $q^2$ for $B^\pm\to\pi^\pm\ell^+\ell^-$
gets diluted.
But the LHCb experiment could target the low $q^2$ region,
even below 2 GeV$^2$, to zoom in on the effect.
%

The CP asymmetry for $B^0\to\pi^0\ell^+\ell^-$ is small,
as ``weak annihilation'' is mediated by
color-suppressed $W$-exchange process $\bar b d \to \bar uu$
or loop-suppressed QCD penguins,
while the charge of the spectator quark
leads to a factor of 2 reduction.
Measuring this small asymmetry would be challenging, 
even for the Belle II experiment.

The direct CP asymmetries for $B\to K\ell^+\ell^-$ can arise via similar mechanism,
but is highly suppressed by the hierarchy of CKM factors,
$|V_{us}^*V_{ub}| \ll |V_{ts}^*V_{tb}|$.
This is consistent with non-observation of CP asymmetry in this mode
reported recently by LHCb~\cite{Aaij:2013dgw}.

The $q^2$-dependence of the isospin asymmetry $A_{\rm I}$ is
given in Fig.~\ref{fig:asym}.
While it gets larger at lower $q^2$,
the asymmetry is far below 10\%.
$A_{\rm I}$ can also be generated by hard-spectator scatterings,
as in the case for CP asymmetries.
However, the effect is reduced since we
take CP average in Eq.~(\ref{def:AI}),
which smooths the oppositely varying
$B^+$ and $B^-$ decay rates at lower $q^2$.
We obtain the $q^2$-averaged isospin asymmetry in our range as
\begin{align}
\la A_{\rm I}\ra
&=0.03^{+0.01}_{-0.00}\;\big|_{\mathrm{CKM}}\,{}^{+0.01}_{-0.01}\big|_{\mathrm{had.}}
\,{}^{+0.00}_{-0.00}\big|_{\mu} \notag\\
&=0.03 \pm 0.01.
\end{align}
%
Measurement of the isospin asymmetry depends on
measuring $B^0\to\pi^0\ell^+\ell^-$,
but appears rather difficult.


%

%
\begin{table}[t!]
\caption[]{
SM predictions in three different $q^2$-ranges
for improved branching ratios (in units of $10^{-8}$),
$q^2$-averaged CP and isospin asymmetries.
The $R_+$-related quantities $F_+^2$, $c_+$ and $d_+$
(Eq.~(\ref{eq:Rplus}))
are also given.
Only total errors are shown.
}
{
$$
\begin{array}{c|ccc}
\hline\hline
 &(2,6)~{\rm GeV}^2 & (1,6)~{\rm GeV}^2 & (1,8)~{\rm GeV}^2 \\
\hline
\hat{\mathcal B}(B^+\to\pi^+\ell\ell) & 0.47^{+0.06}_{-0.04}
 & 0.61^{+0.07}_{-0.06} & 0.82^{+0.10}_{-0.07} \\
\hat{\mathcal B}(B^-\to\pi^-\ell\ell) & 0.36^{+0.05}_{-0.04}
 &  0.44^{+0.06}_{-0.05} & 0.63^{+0.09}_{-0.07} \\
\hat{\mathcal B}(B^0\to\pi^0\ell\ell) & 0.19\pm 0.02
 &0.24^{+0.03}_{-0.02} &0.34^{+0.04}_{-0.03} \\
\hat{\mathcal B}(\bar B^0 \to\pi^0\ell\ell) & 0.18\pm 0.02
 &0.23^{+0.03}_{-0.02} & 0.32^{+0.05}_{-0.04}\\
\hline
\langle A_{\rm CP}^+\rangle & -0.13^{+0.02}_{-0.03}
 &-0.16\pm 0.02 &-0.13\pm 0.02 \\
\langle A_{\rm CP}^0\rangle & -0.03^{+0.01}_{-0.02}
  &-0.03^{+0.01}_{-0.02} &-0.03^{+0.01}_{-0.02} \\
\langle A_{\rm I}\rangle & 0.03\pm 0.01
 & 0.04\pm 0.01 &0.03\pm 0.01 \\
\hline
F_+^2 & 0.58^{+0.09}_{-0.08} &0.59^{+0.09}_{-0.08} &0.60^{+0.09}_{-0.08} \\
c_+ & 0.25^{+0.07}_{-0.06} &0.21^{+0.08}_{-0.07}  & 0.20^{+0.06}_{-0.05}  \\
d_+ & 0.13^{+0.04}_{-0.03} & 0.20\pm 0.06  &  0.15^{+0.04}_{-0.03}\\
\hline\hline
\end{array}
$$
}
\label{tab:obs-app}
\end{table}
\begin{figure*}[t!]
\begin{center}
{\includegraphics[width=70mm]{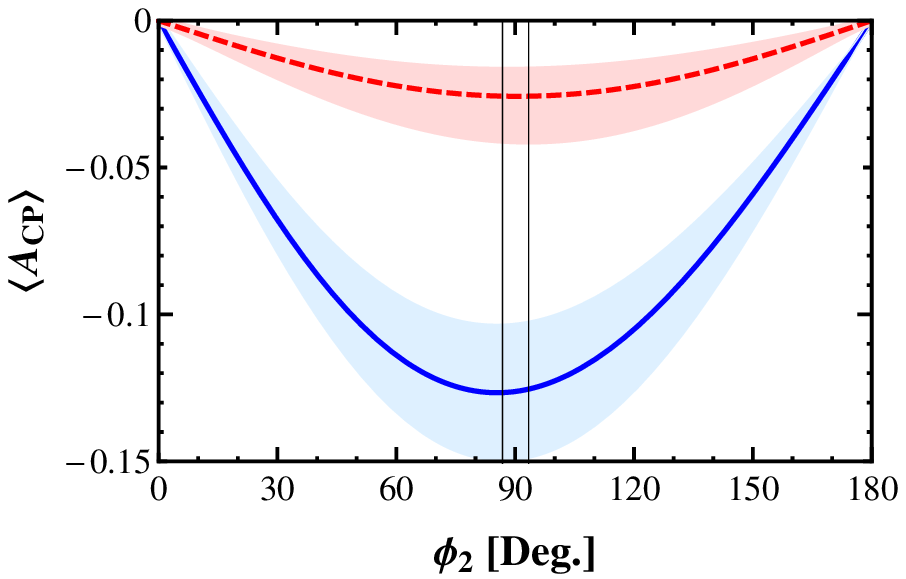} \hskip0.5cm
 \includegraphics[width=70mm]{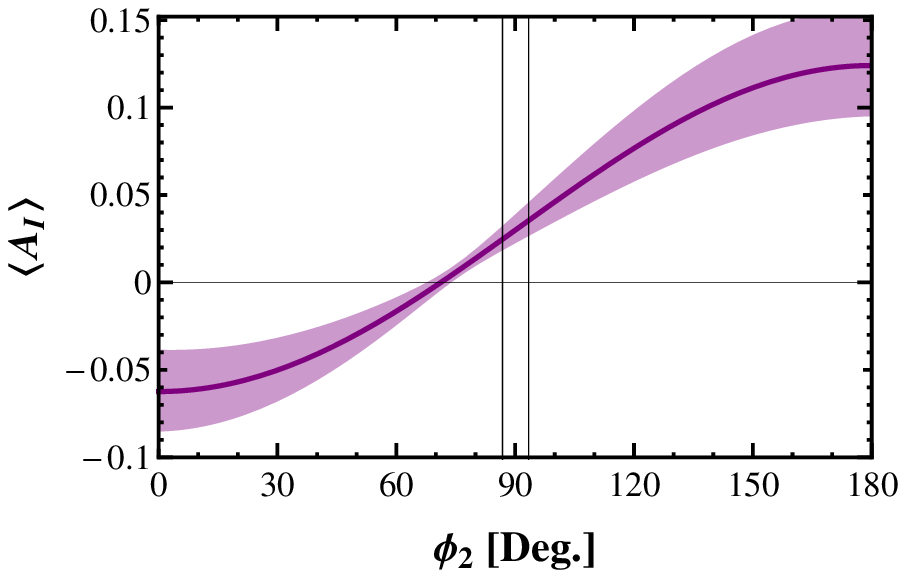}
}
\end{center}
\vskip-0.35cm
\caption{
The $\phi_2$ dependence of $q^2$-averaged asymmetries for
$B\to\pi\ell^+\ell^-$ with $R_{ut}=0.39$:
[left] direct CP asymmetries for $B^+\to \pi^+\ell^+\ell^-$ (blue-solid)
and $B^0\to \pi^0\ell^+\ell^-$ (red-dashed);
[right] isospin asymmetry for $B\to \pi\ell^+\ell^-$ decays.
Shaded regions indicate the combined hadronic and
scale uncertainties, 
and black solid lines indicate the global fit~\cite{PDG} value
$\phi_2 = (89^{+4}_{-2})^\circ$.
} \label{fig:phi2}
\end{figure*}
%

So far we have adopted $q^2\in (2,6)~{\rm GeV}^2$ as the range of the $q^2$ integral,
to keep theoretical control.
However, uses of wider $q^2$ ranges can be found in literature,
as mentioned in Sec.~\ref{sec:Formula}, and our choice may turn out to be too
conservative after experimental data or more model-independent studies on 
the resonant regions emerge.
Given this,
we also provide in Table~\ref{tab:obs-app} predictions
for various observables for other two wider $q^2$ ranges, namely,
$q^2\in (1,6)~{\rm GeV}^2$ and $(1,8)~{\rm GeV}^2$,
side by side with our adopted range of $q^2\in (2,6) ~{\rm GeV}^2$.
We note that our error estimate does not take into account $1/m_b$
corrections nor systematic errors due to use of model
functions for the $B$-meson light-cone distribution amplitudes.
Some discussion on model dependence is deferred to
Appendix~\ref{app:model}.

\section{\label{sec:CKM} \boldmath
Constraining CKM Parameters \protect\\}

So far we have assumed the Wolfenstein parameters
determined by global fit,
listed in Table~\ref{tab:input-para}, as CKM inputs.
However, once the $B\to \pi\ell^+\ell^-$ observables are
measured with some accuracy,
one may use the measured values to determine the CKM parameters
assuming SM.
In particular, the $B\to \pi\ell^+\ell^-$ observables are sensitive to
$\phi_2\equiv\alpha$, an angle in the $b\to d$ unitarity triangle.

We show in Fig.~\ref{fig:phi2} the $\phi_2$ dependence
of the $q^2$-averaged asymmetries with fixed
$R_{ut} = |\lambda_u/\lambda_t| = 0.39$
(see Eq.~(\ref{eq:Rut}) for definition).
For $\phi_2=(89^{+4}_{-2})^\circ$ from global analysis~\cite{PDG},
the direct CP asymmetries are close to maximum
for both neutral and charged $B$ decays.
On the other hand,
the isospin asymmetry for this $\phi_2$ value is small,
located near the vanishing point.



LHCb has also reported~\cite{LHCb:2012de}
the ratio of $B^+\to\pi^+\mu^+\mu^-$ and $B^+\to K^+\mu^+\mu^-$
branching ratios,
\begin{align}
&
\frac{\mathcal B(B^+\to \pi^+\mu^+\mu^-)}{\mathcal B(B^+\to K^+\mu^+\mu^-)}
\notag\\
&= 0.053\pm0.014\; ({\rm stat.}) \pm0.001\;({\rm syst.}).
\end{align}
LHCb utilized this to determine the ratio $|V_{td}|/|V_{ts}|$,
with form factors and Wilson coefficients as theoretical input.
As the formulas in Sec. II can be straightforwardly applied to
$B^+\to K^+\ell^+\ell^-$ decay,
one can calculate the corresponding quantity in low $q^2$ region
based on QCDF.

For our estimate, we define
\begin{align}
R_+
&\equiv
\frac{\overline{\mathcal{B}}(B^+\to \pi^+\ell^+\ell^-)}
{\overline{\mathcal{B}}(B^+\to K^+\ell^+\ell^-)},
\end{align}
where $\overline{\mathcal{B}}$ are CP-averaged branching ratios
integrated over $2$ GeV$^2 < q^2 < 6$ GeV$^2$.
Besides overall $|V_{td}|/|V_{ts}|$ dependence,
$R_+$ also depends on $R_{ut}$ and $\phi_2$ through
$\overline{\mathcal{B}}(B^+\to\pi^+\ell^+\ell^-)$.
More explicitly, CKM dependence can be extracted from
\begin{align}
R_+
&= \left|\frac{V_{td}}{V_{ts}}\right|^2 F_+^2
 \left[ 1- c_+ R_{ut}\cos\phi_2 +d_+R_{ut}^2 \right],
\label{eq:Rplus}
\end{align}
which involves the CKM-independent quantities,
\begin{align}
F_+^2
&= \left( \frac{\xi_\pi(0)}{\xi_K(0)} \right)^2
 \frac{ \langle |\mathcal C_{9,\pi^-}^{(t)}|^2 +C_{10}^2 \rangle_{\pi^-} }
 { \langle |\mathcal C_{9,K^-}^{(t)}|^2 +C_{10}^2 \rangle_{K^-} }, 
\label{eq:F+}
\end{align}
and
\begin{align}
c_+
&= 2\frac{\langle{\rm Re}[\mathcal C_{9,\pi^-}^{(t)*}\mathcal C_{9,\pi^-}^{(u)}] \rangle_{\pi^-}}
 {\langle |\mathcal C_{9,\pi^-}^{(t)}|^2 +C_{10}^2 \rangle_{\pi^-}}, 
\ \
d_+
= \frac{\langle |\mathcal C_{9,\pi^-}^{(u)}|^2 \rangle_{\pi^-}}
 {\langle |\mathcal C_{9,\pi^-}^{(t)}|^2 +C_{10}^2 \rangle_{\pi^-}}.
\label{eq:c+d+}
\end{align}
We have introduced, for a $q^2$-dependent quantity $X(q^2)$,
an abbreviation
\begin{align}
\langle X \rangle_P
\equiv \int_{2~{\rm GeV}^2}^{6~{\rm GeV}^2}dq^2 \lambda(q^2,m_P^2)^3
 \left( \frac{\xi_P(q^2)}{\xi_P(0)} \right)^2 X(q^2).
\end{align}
$F_+^2$ corresponds to $f^2$ in Ref.~\cite{LHCb:2012de} with $R_{ut}=0$,
while $c_+$ and $d_+$ deform the correspondence when $R_{ut}\neq 0$.

Our numerical result for $2~\rm{GeV}^2 < q^2 < 6~\rm{GeV}^2$
(results for other $q^2$ ranges are given in Table~\ref{tab:obs-app}) is
\begin{align}
&F_+^2=0.58_{-0.08}^{+0.09}, \ \
c_+=0.25_{-0.06}^{+0.07}, \ \ d_+=0.13_{-0.03}^{+0.04}.
\end{align}
For $R_{ut} = 0.39$ and $\phi_2=89^\circ$ from global analysis~\cite{PDG},
the terms including $c_+$ or $d_+$ in Eq.~(\ref{eq:Rplus}) contribute by
only a few percent.
In this case, the main theoretical uncertainty in Eq.~(\ref{eq:Rplus})
comes from the overall $F_+^2$, where the uncertainty is dominated
by the one from the form factor ratio $\xi_K(0)/\xi_\pi(0)$.
The QCD light-cone sum rule result~\cite{Duplancic:2008tk},
which takes the SU(3) breaking corrections into account,
provides this form factor ratio with better precision
than the individual ones (see Table \ref{tab:input-para}).
Therefore, the error of $F_+^2$, which is around 15\%,
is much smaller than the 30 or 20\% error in the original $B^+\to \pi^+\ell^+\ell^-$
branching ratio,
shown in Eq.~(\ref{eq:BR-num}).

Of course, our result can not be directly compared with the LHCb result,
where the full $q^2$ range seems to be used.
However, it should become possible once statistics is increased in the near future.
Although the ratio $\Delta m_{B_d}/\Delta m_{B_s}$ is rather well measured,
this method can provide a complementary check.
Once several $B\to \pi\ell^+\ell^-$ observables are measured with certain precision,
their combination would be useful to constrain the CKM parameters,
adding further information for existing studies on the unitarity triangle.
In the remaining part of this section, we illustrate
the possible impact of future measurements of these observables.

\begin{figure}[t!]
\begin{center}
{
\includegraphics[width=7.2cm]{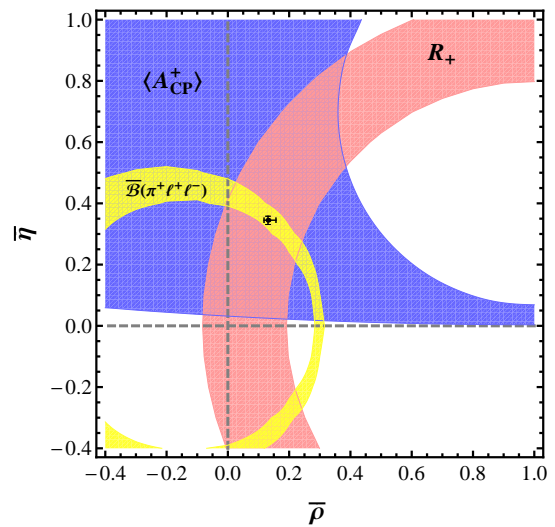}
}
\end{center}
\caption{
Possible constraints on $(\bar\rho,\;\bar\eta)$ plane from
CP-averaged $\overline{\mathcal B}(B^+\to\pi^+\ell^+\ell^-)$
(improved via Eq.~(\ref{eq:decayrate-imp})),
$R_+$ and $\langle A_{\rm CP}^+\rangle$,
assuming the guesstimates
given in Eqs.~(\ref{projection-BR}) and (\ref{projection-ACP}).
The allowed regions are drawn by also taking into account
theoretical uncertainty (except CKM) for these observables,
added linearly with experimental errors.
The black dot with error bars indicates preference of global analysis~\cite{PDG}.
}
\label{fig:rho-eta}
\end{figure}

Despite the observation of the $B^+\to \pi^+\mu^+\mu^-$ mode
by LHCb with $1.0$ fb$^{-1}$ data~\cite{LHCb:2012de},
it might take a while to establish its branching ratio in our $q^2$ range,
$2~{\rm GeV}^2 < q^2 < 6~{\rm GeV}^2$, due to limited statistics.
Nevertheless, LHCb will update with full 2011--2012 data,
and with Run 2 of the LHC at 13--14 TeV to start in 2015,
precise measurements of the $B^+\to\pi^+\ell^+\ell^-$
branching ratio in our $q^2$ range may become possible.
With this anticipation, we make the following guesstimate
for the CP-averaged branching ratio
$\overline{\mathcal B}(B^+\to\pi^+\ell^+\ell^-)$
and $R_+$ in our $q^2$ range, for LHCb with Run 2 data:
\begin{align}
&\overline{\mathcal B}(B^+\to\pi^+\ell^+\ell^-)^{\rm exp}
=(0.42\pm 0.04)\times 10^{-8}, \notag\\
&R_+^{\rm exp}
=0.027\pm 0.003, \label{projection-BR}
\end{align}
where we take 10\% experimental error with
SM-like central values.
At this precision for the branching ratio,
the SM-like value for the $q^2$-averaged
$\langle A_{\rm CP}^+\rangle \sim -10\%$
would still be difficult to establish.
We simply assume
\begin{align}
\langle A_{\rm CP}^+\rangle^{\rm exp}
= -0.13\pm 0.10,
 \label{projection-ACP}
\end{align}
but refinement, especially zooming in on larger asymmetry
at lower $q^2$, can be done by LHCb.

Possible constraints on the $(\bar\rho,\;\bar\eta)$ plane with these projections
for future measurements are illustrated in Fig.~\ref{fig:rho-eta}.
$\overline{\mathcal B}(B^+\to\pi^+\ell^+\ell^-)$
 [improved via Eq.~(\ref{eq:decayrate-imp})]
provides a narrower allowed ring than $R_+$,
because of smaller theoretical uncertainty
as well as stronger dependence on $\bar\rho$ and $\bar\eta$,
brought about by $V_{ub}=A\lambda^3(\rho-i\eta)$
from $B\to\pi\ell\nu$ rate.
One notes that the combined constraint of $\overline{\mathcal B}(B^+\to\pi^+\ell^+\ell^-)$
and $R_+$ allows two strip-shaped regions.
If $\langle A_{\rm CP}^+\rangle$ is found with
negative sign, as in Eq.~(\ref{projection-ACP}),
it would favor the upper strip.
Note that $\langle A_{\rm CP}^+\rangle$ changes sign
around $\bar\eta=0$:
$\langle A_{\rm CP}^+\rangle < 0$ for $\bar\eta > 0$
and vice versa.
Therefore, even though establishing the value of
$\langle A_{\rm CP}^+\rangle$ in the near future may be challenging,
measuring the sign of $\langle A_{\rm CP}^+\rangle$
would still provide useful information to
discriminate between the two separate regions.

For the neutral $B$ decay case,
so far only upper limits are available (see Table~\ref{tab:exp}).
The Belle II experiment may improve the situation in the future.
But judging from the poor performance of $B^0\to \pi^0 \ell^+\ell^-$
by Belle, as indicated in Table~\ref{tab:exp},
there is some worry whether this rate can be measured well for Belle II.
The situation for $\langle A_{\rm CP}^0\rangle$
and $\langle A_{\rm I}\rangle$ would be much worse
due to the small SM predictions.
%
However, even if measurements for $B^0\to \pi^0\ell^+\ell^-$
is not forthcoming, the charged $B$ decay observables by themselves
can still provide useful information, as we have explained.
We do not show possible constraints by neutral $B$ decay related
observables.\footnote{
We have checked that constraints from
$\overline{\mathcal B}(B^0\to\pi^0\ell^+\ell^-)$
assuming 10\% experimental error provide
similar allowed region as
$\overline{\mathcal B}(B^+\to\pi^+\ell^+\ell^-)$
in Fig.~\ref{fig:rho-eta}.
Such measurements, however,
do not seem reachable in the foreseeable future.}


%

\section{\label{sec:Discussion} \boldmath
Discussion and Conclusion\protect\\}

Observation of the rare decay $B^+\to \pi^+\mu^+\mu^-$ indicates that  the
$b\to d\ell^+\ell^-$  precision era has dawned.
As a preparation for the advent of more precise measurements,
our systematic study of the semileptonic decay $B\to \pi\ell^+\ell^-$
based on the QCDF framework suggests
rich information, and sizable direct CP asymmetry is
predicted for charged $B$ decay for $q^2 \lesssim 3$--4 GeV$^2$.
Such large CP asymmetry could be background of
the CP asymmetry measurement in $B^+\to K^+\mu^+\mu^-$ decays,
which seems to be glossed over in the LHCb study~\cite{Aaij:2013dgw}.

We advocate the $q^2$ region between 2 to 6 GeV$^2$,
where QCDF is applicable.
Hadronic uncertainty due to form factors are further reduced
by taking ratio with $B^0 \to \pi^-\ell^+\nu$ rate.
We have shown, due to large absorptive part arising
from hard-spectator-scattering, i.e. the annihilation diagram of
Fig.~\ref{fig:WeakAnnih}, together with the CP phase via $V_{ub}$ and $V_{td}$,
leads to a sizable CP asymmetry $A_{\rm CP}^+$ in
decay rate of $B^+ \to \pi^+\ell^+\ell^-$,
which grows for lower $q^2$.
Similar effect is suppressed for $B^0 \to \pi^0\ell^+\ell^-$,
which has a rather small direct CP asymmetry.
We also study the isospin asymmetry $A_{\rm I}$
between the charged and neutral $B$ decays.
While it also grows with lower $q^2$,
the effect remains below 10\% and smaller than $A_{\rm CP}^+$.
Given that it would require the
measurement of $B^0 \to \pi^0\ell^+\ell^-$,
which can only be done by Belle II,
its measurement is more distant in the future.

To emphasize the importance and utility of further,
improved measurements of $B^+ \to \pi^+\ell^+\ell^-$,
we illustrate its potential for crosschecking the
three generation SM,
by showing the intersect of the allowed region of
$B^+ \to \pi^+\ell^+\ell^-$ rate normalized by
$B \to \pi \ell\nu$, as well as normalized by
$B^+ \to K^+\ell^+\ell^-$, which bring in
different CKM dependence.
Assuming some future precision with the SM-like central values,
we show that it has two intersecting regions.
Interestingly, by measuring the {\it sign} of
$A_{\rm CP}^+$ (rather than a more precise value measurement),
one can eliminate one of the regions,
hence provide a measure on the $(\bar\rho,\;\bar\eta)$ plane.
Though not spectacular, it provides independent
information from, e.g. $\Delta m_{B_d}/\Delta m_{B_s}$.

The direct CP asymmetry gets enhanced at low $q^2$,
a firm consequence of the annihilation diagram that
participates to $B^+ \to \pi^+\ell^+\ell^-$ at tree level.
Since the trend continues as $q^2$ gets lower, we have
explored below 2 GeV$^2$, beyond the
region of applicability for QCDF,
and entering the hadronic resonance region.
With increased data, LHCb should target this low
$q^2$ region, below 3--4 GeV$^2$, to check this
enhancement. This not only facilitates the
CKM measurement program, but also checks the presence
of the annihilation diagram.

The annihilation diagram provides a tree level mechanism
for bringing in a large strong phase
through the on-shell $u$ quark from the $B^+$ wavefunction
(or light-cone distribution amplitude),
by emitting a photon that becomes an $\ell^+\ell^-$ pair.
The strength of $A_{\rm CP}^+$ reaches beyond $-25\%$ for
$q^2$ around 2 GeV$^2$, and the trend continues as
$q^2$ is lowered.
Recently, the LHCb experiment has measured CP asymmetries
in 3-body $B^\pm \to K^\pm h^+h^-$~\cite{Khh-LHCb} and
$B^\pm \to \pi^\pm h^+h^-$~\cite{pihh-LHCb} decays,
where $h^+h^-$ stands for $\pi^+\pi^-$ or $K^+K^-$.
All modes give large ``regional'' CP asymmetries,
with absolute values reaching beyond 50\%, except the $K^\pm K^+K^-$ final state.
The CP asymmetry Dalitz plots for $B^\pm \to K^\pm h^+h^-$
and $\pi^\pm \pi^+\pi^-$ exhibit complex structures, and
contain a lot of hadronic effects.
Of particular interest to us is $B^\pm \to \pi^\pm K^+K^-$,
where LHCb finds CP asymmetry as large as $\sim -60\%$
for $m^2_{K^+K^-} < 1.5$ GeV$^2$, which is
reminiscent of our finding for $B^\pm \to \pi^\pm \ell^+\ell^-$ decay.
Indeed, replacing $K^+K^-$ by $\ell^+\ell^-$ gives
the $B^\pm \to \pi^\pm \ell^+\ell^-$ process.
Could the observed effect occur through a similar mechanism
as we discussed? It certainly cannot occur through
Fig.~\ref{fig:WeakAnnih} with a virtual photon,
as $\alpha$ is way too weak compared to hadronic couplings.
However, we speculate that, if the virtual photon is
replaced by some hadron that couples to both
$u\bar u$ and $s\bar s$, a similar effect should be
achievable, and the absorptive part can come from
$u$ quark being on-shell in Fig.~\ref{fig:WeakAnnih},
rather than from the resonance width.
We will return to a model study of the $B^\pm \to \pi^\pm K^+K^-$
for low $m^2_{K^+K^-}$ in a future work.

In conclusion, we have studied $B^{+,0}\to \pi^{+,0}\ell^+\ell^-$ decays
in the framework of QCDF, giving the SM prediction
for the branching ratios in the $q^2 = m^2_{\ell^+\ell^-}$ range of 2 to 6 GeV$^2$,
together with a slightly larger range for comparison.
We study also the associated CP and isospin asymmetries.
A relatively precise prediction is made for
${\cal B}(B^{+}\to \pi^{+}\ell^+\ell^-)$ that can be tested by
experiment. Furthermore, we find that, due to annihilation diagram
with hard-spectator scattering,
the CP asymmetry grows with lower $q^2$ and can become rather sizable,
which is another feature that can be tested by experiment.
Future precision measurements can provide a check on CKM parameters.
Our mechanism for large CP asymmetries at low $q^2$ seems to
echo the large CP asymmetry for low $m^2_{K^+K^-}$
observed by LHCb for $B^{\pm}\to \pi^{\pm}K^+K^-$.

\vskip0.3cm
\noindent{\bf Acknowledgement}.  WSH is supported by the the
Academic Summit grant NSC 102-2745-M-002-001-ASP of the
National Science Council, as well as by grant NTU-EPR-102R8915.
MK is supported under NTU-ERP-102R7701 and the Laurel program,
and FX under NSC 102-2811-M-002-205. FX especially acknowledges
the hospitality of Shanghai Jiao Tong University, and
Kavli Institute for Theoretical Physics China at the Chinese Academy of
Sciences, in which part of the work was done.

\appendix

\section{\label{app:formula} \boldmath
$\bar B\to P\gamma^*$ amplitude\protect}


As already given in Eq.~(\ref{QCDF-B2Pgam}),
the $\bar B \to P\gamma^*$ amplitude at $\mathcal O(\alpha_s)$
in the heavy quark limit is given by
\begin{align}
&\mathcal T_P^{(i)}
= \xi_P C_P^{(i)} \notag\\
&+\zeta_P \sum_{\pm}
 \int_0^\infty \frac{d\omega}{\omega}\Phi_{B,\pm}(\omega)\int_0^{1}du\,\phi_P(u)
 T_{P,\pm}^{(i)}(u,\omega)
\end{align}
where
\begin{align}
 \zeta_P & = \frac{\pi^2}{N_c}\frac{f_Bf_P}{M_B}, \\
 C_P^{(i)} & = C_P^{(0,i)} +\frac{\alpha_s C_F}{4\pi} C_P^{(1,i)}, \\
 T_{P,\pm}^{(i)}(u,\omega)
 & = T_{P,\pm}^{(0,i)}(u,\omega) +\frac{\alpha_s C_F}{4\pi} T_{P,\pm}^{(1,i)}(u,\omega),
\end{align}
and $i=t,\;u$. One can show that
the $\bar B\to P\gamma^*$ amplitude is related to the $\bar B\to V_{\parallel}\gamma^*$
amplitude, where $V_\parallel$ is longitudinally polarized, through
\begin{align}
C_P^{(i)} &= -C_{\parallel}^{(i)}, \quad
T_{P,\pm}^{(i)}(u,\omega) = -T_{\parallel,\pm}^{(i)}(u,\omega).
\end{align}
Therefore, one can utilize the existing result for $\bar B\to \rho\gamma^*$~
\cite{Beneke:2004dp} to obtain the $\bar B\to \pi\gamma^*$ amplitudes.
The explicit expressions for
the $\bar B \to P\gamma^*$ amplitudes
are
\begin{align}
&\mathcal T_P^{(t)}
 = \xi_P\left( C_P^{(0,t)} +\frac{\alpha_sC_F}{4\pi}\left[ C_P^{(f,t)} +C_P^{(nf,t)} \right] \right)
 \notag\\
 & \ +\zeta_P
  \lambda_{B,-}^{-1} \int du\,\phi_{P}(u) \hat T_{P,-}^{(0,t)} \notag\\
 & \ +\frac{\alpha_sC_F}{4\pi} \zeta_P \bigg(
  \lambda_{B,+}^{-1} \int du\,\phi_{P}(u)\left[ T_{P,+}^{(f,t)}(u) +T_{P,+}^{(nf,t)}(u) \right]\notag\\
 & \quad\quad\quad\quad\ \ \; +\lambda_{B,-}^{-1} \int du\,\phi_{P}(u) \hat T_{P,-}^{(nf,t)}(u) \bigg),
\label{eq:amp-t}
\end{align}
\begin{align}
& \mathcal T_P^{(u)}
=\ \xi_P\left( C_P^{(0,u)} +\frac{\alpha_sC_F}{4\pi}C_P^{(nf,u)} \right) \notag\\
 & \ + \zeta_P
  \lambda_{B,-}^{-1} \int du\,\phi_P(u) \hat T_{P,-}^{(0,u)} \notag\\
 & \ + \frac{\alpha_sC_F}{4\pi} \zeta_P \bigg(
  \lambda_{B,+}^{-1} \int du\,\phi_P(u) T_{P,+}^{(nf,u)}(u) \notag\\
 & \quad\quad\quad\quad\ \ \; + \lambda_{B,-}^{-1} \int du\,\phi_P(u) \hat T_{P,-}^{(nf,u)}(u) \bigg),
\label{eq:amp-u}
\end{align}
where we have introduced
\begin{align}
\hat T^{(0,i)}_{P,-}
&\equiv \frac{M_B\omega-q^2-i\epsilon}{M_B\omega}T^{(0,i)}_{P,-}(u,\omega), \notag\\
\hat T^{(nf,i)}_{P,-}(u)
&\equiv \frac{M_B\omega-q^2-i\epsilon}{M_B\omega}T^{(nf,i)}_{P,-}(u,\omega),
\end{align}
to remove $\omega$ dependence.

The $\mathcal O(\alpha_s^0)$ contributions to the form factor term are
\begin{align}
C_P^{(0,t)}
 &= C_7^{\rm eff} +\frac{M_B}{2m_b}Y(q^2),\notag \\
C_P^{(0,u)}
 &= \frac{M_B}{2m_b}Y^{(u)}(q^2),
\end{align}
while the $\mathcal O(\alpha_s)$ factorizable and nonfactorizable corrections
are ($C_F=4/3$)
\begin{align}
C_P^{(f,t)}
&=\ \left( \ln\frac{m_b^2}{\mu^2} +2L +\Delta M \right) C_7^{\rm eff}, \\
C_F C_P^{(nf,t)}& =\ -\overline C_2F_2^{(7)} -C_8^{\rm eff}F_8^{(7)} \notag\\
 -\ & \frac{M_B}{2m_b} \bigg[ \overline C_2F_2^{(9)}
  +2\overline C_1 \bigg(F_1^{(9)}+\frac{1}{6}F_2^{(9)}  \bigg) \notag\\
&\ \ \ \ + C_8^{\rm eff}F_8^{(9)} \bigg], \\
C_F C_P^{(nf,u)}
& =\ -\;\overline C_2 \left(F_2^{(7)}+F_{2,u}^{(7)}\right) \notag\\
-\ &\frac{M_B}{2m_b}\bigg[ \overline C_2\left(F_2^{(9)}+F_{2,u}^{(9)}\right) \notag\\
 +\ &2\overline C_1\left(\left(F_1^{(9)}+F_{1,u}^{(9)}\right)
  + \frac{1}{6}\left(F_2^{(9)}+F_{2,u}^{(9)}\right) \right)
 \bigg].
\end{align}

\begin{figure*}[t!]
{\includegraphics[width=70mm]{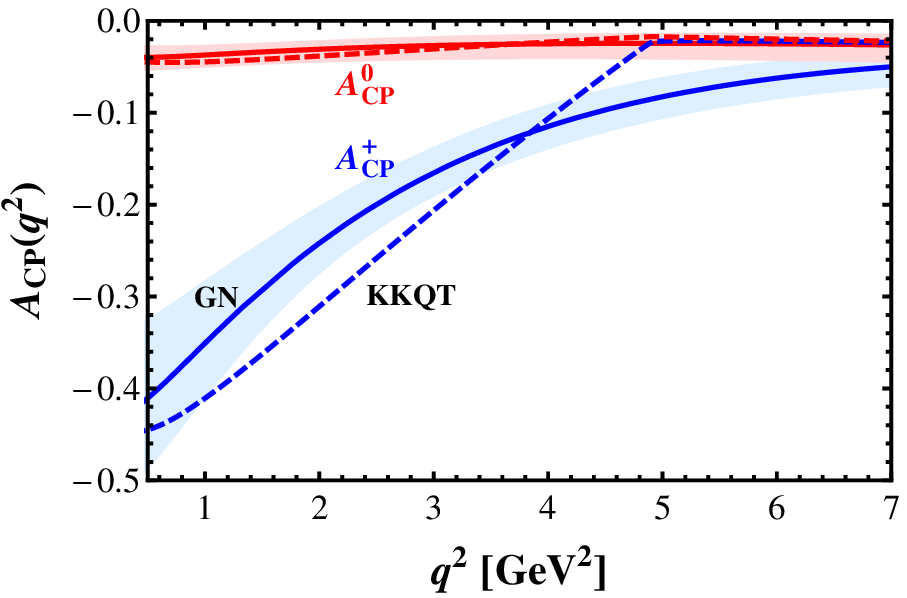} \hskip0.5cm
 \includegraphics[width=70mm]{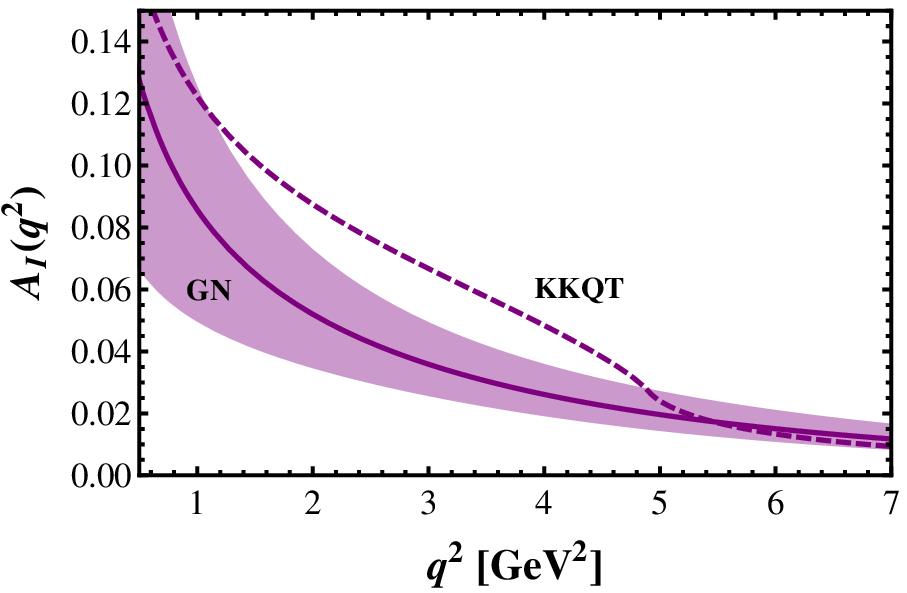}
}
\vskip-0.35cm
\caption{
$A_{\rm CP}(q^2)$ [left] and $A_{\rm I}(q^2)$ [right] obtained by using
``KKQT'' $B$-meson light-cone distribution amplitudes~\cite{Kawamura:2001jm}
(dashed),
compared with the ``GN'' model (solid) result given in the right panel of Fig.~\ref{fig:B-dif}
and Fig.~\ref{fig:asym},
where shaded bands denote theoretical uncertainty of the GN model result.
} \label{fig:KKQT}
\end{figure*}

The $\mathcal O(\alpha_s^0)$ contributions to hard-spectator-scattering term
from weak annihilation diagrams of Fig.~\ref{fig:WeakAnnih} are
\begin{align}
&\hat T^{(0,t)}_{P,-} = e_q\frac{4M_B}{m_b}C_q^{34},\;\;
\hat T^{(0,u)}_{P,-} = -e_q\frac{4M_B}{m_b}C_q^{12},
\label{eq:T0}
\end{align}
where $q=u,d$ is the flavor of the spectator quark in the $B$ meson
with $e_q$ ($q=u,d$) its charge, and
\begin{align}
C_q^{34}
&\equiv C_3+\frac{4}{3}(C_4+12C_5+16C_6), \\
C_q^{12}
&\equiv 3\,\delta_{qu}\,C_2 -\delta_{qd}\left( \frac{4}{3}C_1 +C_2\right).
\end{align}
The $\mathcal O(\alpha_s)$ factorizable and nonfactorizable corrections to
the hard-spectator-scattering term are given by
\begin{align}
T_{P,+}^{(f,t)}(u)
 &= -C_7^{\rm eff}\frac{4M_B}{\bar u E}, \\
T_{P, +}^{(nf,t)}(u)
 &= -\frac{M_B}{m_b} \big[ e_ut_\parallel(u,m_c)(\overline C_2+ \overline C_4
  -\overline C_6)
 \notag\\
 +\ & e_dt_\parallel(u,m_b)(\overline C_3+ \overline C_4 -\overline C_6)
 +e_dt_\parallel(u,0)\overline C_3 \big], \\
T_{P, +}^{(nf,u)}(u)
 &= -e_u\frac{M_B}{m_b}\left( C_2-\frac{1}{6}C_1 \right) \notag\\
& \quad\quad\quad \times \left[ t_\parallel(u,m_c) -t_\parallel(u,0) \right], \\
\hat T_{P, -}^{(nf,t)}(u)
&= -e_q \bigg[ \frac{8C_8^{\rm eff}}{\bar u +uq^2/M_B^2}
 \notag\\
 & + \frac{6M_B}{m_b} \Big\{ h(\bar uM_B^2+uq^2,m_c)(\overline C_2
  +\overline C_4+\overline C_6) \notag\\
 & \quad\quad + h(\bar uM_B^2+uq^2,m_b)(\overline C_3+\overline C_4+\overline C_6)
 \notag\\
 & \quad\quad + h(\bar uM_B^2+uq^2, 0)(\overline C_3+3\overline C_4+3\overline C_6)
 \notag \\
 & \quad\quad -\frac{8}{27}(\overline C_3 -\overline C_5 -15\overline C_6)
 \Big\} \bigg], \\
\hat T_{P, -}^{(nf,u)}(u)
&= -e_q \frac{6M_B}{m_b}\left( C_2-\frac{1}{6}C_1 \right) \notag\\
 & \times \left[ h(\bar uM_B^2+uq^2,m_c)-h(\bar uM_B^2+uq^2, 0) \right],
\end{align}
where $\overline C_i$ (for $i=1,...,6$)
are defined by
\begin{align}
\overline C_1 &= \frac{1}{2}C_1, \quad
\overline C_2 = C_2 -\frac{1}{6}C_1, \notag\\
\overline C_3 &= C_3 -\frac{1}{6}C_4+16C_5-\frac{8}{3}C_6, \quad
\overline C_4 = \frac{1}{2}C_4+8C_6, \notag\\
\overline C_5 &= C_3 -\frac{1}{6}C_4+4C_5-\frac{2}{3}C_6, \quad
\overline C_6 = \frac{1}{2}C_4+2C_6.
\end{align}
Details of the other definitions can be found in
Refs.~\cite{Beneke:2001at,Beneke:2004dp},
and references therein.

\section{\label{app:model} \boldmath
Hadronic model dependence \protect}

We comment that the larger error at lower $q^2$
for asymmetries $A_{\rm CP}$ and $A_{\rm I}$ is
due to the dominance of weak annihilation.
As the latter is calculated at leading order in $1/m_b$ \cite{Beneke:2001at},
there might be $1/m_b$ corrections even at $\mathcal O(\alpha_s^0)$,
whose uncertainty is not included in our error estimate.
Besides, the use of model functions for $B$-meson light-cone
distribution amplitudes might introduce extra uncertainty,
which we will turn to shortly.
For the pion light-cone distribution amplitude $\phi_\pi$,
its uncertainty has little effect on the hadronic errors,
although the Gegenbauer coefficients $a_{2,4}^\pi$ listed in
Table \ref{tab:input-para} are rather uncertain.
This is because the leading order weak annihilation $\hat T_{P,-}^{(0,i)}$
that enters Eqs.~(\ref{eq:amp-t}) and (\ref{eq:amp-u}) does not depend on $u$.
Thus, $\phi_\pi(u)$ can be integrated out, giving simply
$\int du\,\phi_\pi=1$ due to normalization.
Therefore, there is no dependence on the Gegenbauer coefficients
in the leading order weak annihilation terms.
To see the impact of shape functions for
$B$-meson light-cone amplitudes,
we reevaluate previous results using
\begin{align}
\Phi_{B,+}^{\rm KKQT}(\omega)
&= \frac{\omega}{2\bar\Lambda^2}\theta(2\bar\Lambda-\omega), \notag\\
\Phi_{B,-}^{\rm KKQT}(\omega)
&= \frac{2\bar\Lambda-\omega}{2\bar\Lambda^2}\theta(2\bar\Lambda-\omega),
\label{eq:BLDA-KKQT}
\end{align}
from Ref.~\cite{Kawamura:2001jm} (KKQT),
where $\bar\Lambda = M_B-m_b$, and
we use pole mass $m_{b,\rm pole}$
following argument of Ref.~\cite{Neubert:1993mb}.


In our preferred $q^2$ range, modifications
in the $q^2$ distributions $d\hat{\mathcal B}/dq^2$
by the use of these light-cone distribution amplitudes
are small; within 10\% (1\%) for the charged (neutral) $B$ decays.
These shifts are within the error bands shown in the left panel of Fig.~\ref{fig:B-dif}.
On the other hand, as shown in the left panel of Fig.~\ref{fig:KKQT},
the distortion of $A_{\rm CP}^+(q^2)$ is more significant,
although the overall behavior is similar to the ``GN model''
(Ref.~\cite{Grozin:1996pq}) used in the main text.
The $q^2$ distribution drops sharply down to (minus) a few percent level at
$q^2= 2\bar\Lambda M_B\sim 5~{\rm GeV}^2$
due to the sharp $\omega =2\bar\Lambda$ cutoff in Eq.~(\ref{eq:BLDA-KKQT}),
in contrast to the smooth damping  in GN model.
The magnitude of $A_{\rm CP}^+(q^2)$ gets enhanced (suppressed)
below (above) $q^2\sim 4~{\rm GeV}^2$ up to 30\% (70\%).
Nevertheless, the $q^2$-averaged $\langle A_{\rm CP}^+\rangle$ does not change
from the GN model case, because of an
accidental cancellation for our particular choice
of $2~{\rm GeV}^2 < q^2 < 6~{\rm GeV}^2$ range.
$A_{\rm CP}^0(q^2)$ also gets distorted, but the shift is small,
within the error band of the GN model prediction. 

As shown in the right panel of Fig.~\ref{fig:KKQT},
$A_{\rm I}(q^2)$ is enhanced
for $q^2$ below $5.4~{\rm GeV}^2$.
The enhancement is above 50\% for
$1.2~{\rm GeV}^2 \lesssim q^2\lesssim 4.8~{\rm GeV}^2$,
with the maximal enhancement $\sim 90$\% at $q^2\sim 3.5~{\rm GeV}^2$.
The $q^2$-average is also enhanced, $\langle A_{\rm I}\rangle^{\rm KKQT}=0.05$,
but is still below 10\%.


\end{document}